\documentclass{vgtc}                          % final (conference style)
\ifpdf%                                % if we use pdflatex
  \pdfoutput=1\relax                   % create PDFs from pdfLaTeX
  \usepackage{graphicx}                % allow us to embed graphics files
  \DeclareGraphicsExtensions{.pdf,.png,.jpg,.jpeg} % for pdflatex we expect .pdf, .png, or .jpg files
\else%                                 % else we use pure latex
  \usepackage{graphicx}                % allow us to embed graphics files
  \DeclareGraphicsExtensions{.eps}     % for pure latex we expect eps files
\fi%

\graphicspath{{figures/}{pictures/}{images/}{./}} % where to search for the images

\usepackage{microtype}                 % use micro-typography (slightly more compact, better to read)
\PassOptionsToPackage{warn}{textcomp}  % to address font issues with \textrightarrow
\usepackage{textcomp}                  % use better special symbols
\usepackage{mathptmx}                  % use matching math font
\usepackage{times}                     % we use Times as the main font
\usepackage{cite}                      % needed to automatically sort the references
\usepackage{tabu}                      % only used for the table example
\usepackage{booktabs}                  % only used for the table example
\usepackage{todonotes}
\usepackage{enumitem}

\usepackage[utf8]{inputenc}
\usepackage{csquotes}
\usepackage{multirow}

\usepackage{caption}
\usepackage{xurl}

\newcommand{\revision}[1]{{\color{black} #1}}

\onlineid{8813}

\vgtccategory{Research}

\vgtcinsertpkg

\title{\revision{Evaluating the Effect of Enhanced Text-Visualization Integration on Combating Misinformation in Data Story}}

\author{Chengbo Zheng\thanks{e-mail: cb.zheng@connect.ust.hk} %
\and Xiaojuan Ma\thanks{e-mail: mxj@cse.ust.hk}}
\affiliation{\scriptsize Hong Kong University of Science and Technology}

\abstract{
Misinformation has disruptive effects on our lives. Many researchers have looked into means to identify and combat misinformation in text or data visualization. However, there is still a lack of understanding of how misinformation can be introduced when text and visualization are combined to tell data stories, not to mention how to improve the lay public's awareness of possible misperceptions about facts in narrative visualization. In this paper, we first analyze where misinformation could possibly be injected into the production-consumption process of data stories through a literature survey. Then, as a first step towards combating misinformation in data stories, we explore possible defensive design methods to enhance the reader's awareness of information misalignment when data facts are scripted and visualized. More specifically, we conduct a between-subjects crowdsourcing study to investigate the impact of two design methods enhancing text-visualization integration, i.e., explanatory annotation and interactive linking, on users' awareness of misinformation in data stories. The study results show that although most participants still can not find misinformation, the two design methods can significantly lower the perceived credibility of the text or visualizations. Our work informs the possibility of fighting an infodemic through defensive design methods.
} % end of abstract

\CCScatlist{
  \CCScatTwelve{Human-centered computing}{Visu\-al\-iza\-tion}{Visualization design and evaluation methods}{}
}

\begin{document}
\newcommand{\review}[1]{{\color{red} #1}}


\maketitle

%% \section{Introduction} %for journal use above \firstsection{..} instead
\section{Introduction}
In recent years, the proliferation of misinformation on online media has become a paramount public concern. 
During the COVID-19 pandemic, misinformation, such as ingesting bleach as a coronavirus treatment, is a serious threat to public health~\cite{roozenbeek2020susceptibility}.
While it is widely known that false or inaccurate information can be disseminated through text and/or images, visualizations appear to be another increasingly popular vehicle for misinformation~\cite{lee2021viral, pandey2015deceptive ,correll2017black}.
Due to its persuasive power~\cite{6876023}, visualization may be an even more dangerous yet harder to detect tool to manipulate people's opinion.
A typical example of misinformation in visualizations is the inverted axis, which breaks the conventional direction of the axis and makes readers receive a reversal message to the facts~\cite{pandey2015deceptive}.
% \cb{A study done by Law et al.\cite{law2020causal} shows that  people show less awareness of misleading claims that accompanied with visualizations than without visualizations.}
% Visualization, as a powerful tool that communicates data with audiences, makes misinformation more convincing considering its persuasive nature
% In previous studies, researchers have investigated different ways that misinformation can exist in visualizations. For example, deceptive visualizations ...; title-visualization misalignment 

\begin{figure}[tb]
 \centering % avoid the use of \begin{center}...\end{center} and use \centering instead (more compact)
 \includegraphics[width=0.95\linewidth]{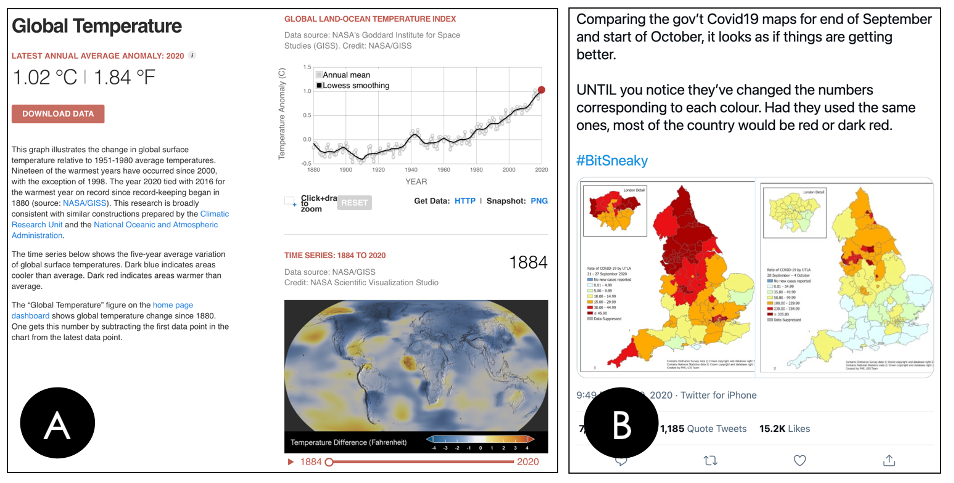}
 \caption{ (A)\cite{nasa_2021}: Narrative visualizations combining rich text and visualizations in online mass media. (B)\cite{andreou_2020}:
Combine text and visualization to tell stories on social media}
 \label{fig:nytimes}
\end{figure}

Prior research has investigated misinformation existed in stand-alone visualization, e.g., bar chart~\cite{pandey2015deceptive, lauer2020people}, thematic maps~\cite{correll2016surprise}, etc.
However, few studies are concerned with how misinformation can be introduced in narrative visualization.
One popular form of narrative visualization as presented in Fig. \ref{fig:nytimes} consists of rich text providing contextual information and some key piece(s) of fact expressed by data visualization~\cite{7274435}.
It is commonly known as illustrated text~\cite{wang2019comparing} (also magazine style~\cite{segel2010narrative}).
While existing research efforts on narrative visualizations mainly concern facilitation of its creation~\cite{shi2020calliope}, there is a pressing demand for in-depth understanding of how problematic design can mislead readers and bring misinformation.
% While existing research efforts on narrative visualization mainly concern facilitation of its creation~\cite{shi2020calliope} , there is a pressing demand for in-depth understanding of how problematic design may mislead readers and bring misinformation.

Moreover, there lacks effective methods that can help lay public stay aware of misinformation in narrative visualizations.
Previous works have explored how to uncover visualization mirage in visual analytics process\cite{10.1145/3313831.3376420} and how to reveal and repair chart errors~\cite{chen2021vizlinter}.
Nevertheless, the interplay between text and visualization in narrative visualizations is rarely considered.
A common approach to combat misinformation is performing automatic or manual fact-checking~\cite{wu2019misinformation}. 
While the number of fact-checking services and algorithms for textual information is growing rapidly~\cite{wu2019misinformation}, not many of them could efficiently and cost-effectively detect misinformation in narrative visualizations. 

% Hence, as an alternative, we are interested in exploring whether there exist some design methods for narrative visualizations that can enhance people's awareness of misinformation in data storytelling.
% This idea is inspired by the framing effect of narrative visualization which shows that different framing methods can result in different interpretation towards the same data story~\cite{6064988}. 

\begin{figure}[tb]
 \centering % avoid the use of \begin{center}...\end{center} and use \centering instead (more compact)
 \includegraphics[width=0.82\linewidth]{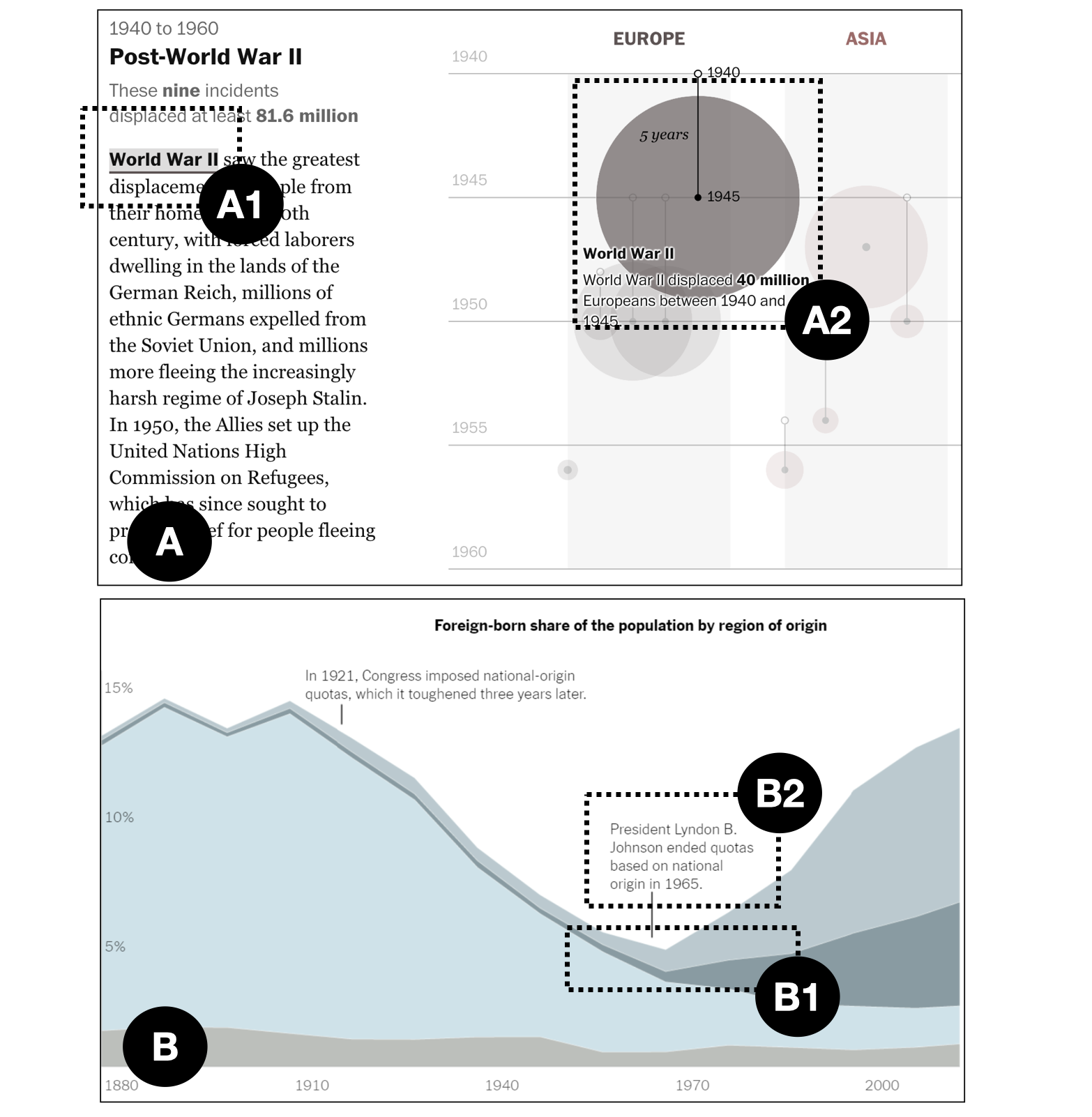}
 \caption{ ((A) Interactive linking\cite{wp_a}: (A1) Highlighted word, (A2)Visualization element corresponding to (A1). When reader hover (A1), (A2) will be highlighted by some visual cues. (B) Explanatory annotation\cite{network_2020}: (B1) Visualization elements of interest, (B2) Text that explain (B1).}
 \label{fig:coneg}
\end{figure}

Facing the aforementioned challenges, in this paper, we begin by asking, ``How might misinformation be introduced into narrative visualization?''.
To this end, we first conducted a literature survey and derived a pipeline that describes the data storytelling production-consumption process.
\revision{
Then, we analyzed every stage of this data storytelling pipeline to identify possible causes of misinformation and then grouped these causes into three categories: author-induced reader-undetectable misinformation (AIRUM), author-induced reader-detectable misinformation (AIRDM) and reader-induced misinformation (RIM).}
% By author-induced misinformation, we refer to false or inaccurate information that is injected by authors in the production process through manipulations invisible to readers.
% By reader-induced misinformation, we refer to misinterpretation that occurs during the consumption process and purely depends on how the visual argument(s) are read in real-world situations.
% By author-reader-co-induced misinformation, we refer to miscommunication that is rooted in the production process but is also affected by readers' comprehension of the output design~\cite{cairo2019charts}.

Next, we proceed to explore possible methods to combat misinformation.
\revision{
Countering AIRUM and RIM might involve moderating the behavior of authors and readers of data stories~\cite{ecker2014correcting, 10.1145/3313831.3376420}, whereas mitigating AIRDM could be achieved by better scrutinizing the presented data stories.
We thus focus on AIRDM in this paper.
Besides, as the first step towards combating misinformation in data stories, we focus on the magazine-style of narrative visualization, which is popular in journalism~\cite{zhi2019linking}.
We target two typical issues for the magazine style under AIRDM : (1) deceptive visualization\cite{pandey2015deceptive}, and (2) text-visualization misalignment\cite{kong2019trust}. 
}

As discussed above, there are still technical barriers to detecting misinformation in data stories.
As an alternative, we are interested in exploring whether there exist some defensive design methods to enhance people's awareness of misinformation, which can prevent readers from falling for problematic data stories in the first place.
This idea is inspired by the fact that different framing of the narrative visualization can result in different interpretations towards the same data story~\cite{6064988}.
Previous studies show that relying on one-sided information to understand data stories may partially result in the unawareness of deceptive visualizations and text-visualization misalignment~\cite{lauer2020people, kong2019trust}.
\revision{
Motivated by this observation, in our study, we considered candidate design methods that can enhance the integration of text and visualization, thus balancing readers' interaction with textual and visual information.
To this end, We choose two commonly used and easy-to-promote design methods: interactive linking and explanatory annotation~\cite{kwon2014visjockey, zhi2019linking, lai2020automatic}.}
% Motivated by this observation, we considered interactive linking and explanatory annotation, the two most common mechanisms to balance the interaction with textual and visual information in journalism~\cite{zhi2019linking, lai2020automatic}, as candidates of defensive design methods to raise readers' awareness of faulty narrative visualization.

Interactive linking (denoted as \textit{Linking} in the rest of the paper) highlights the corresponding explanatory visual elements when selecting specific sentences or words in the text (e.g., Fig.~\ref{fig:coneg}(A)).
Explanatory annotation (denoted as \textit{Annotation}) positions interpretative notes extracted from the text close to the corresponding visual elements~\cite{bryan2016temporal} (e.g., Fig.~\ref{fig:coneg}(B)). Both candidates bridge text and visualization, two critical components to data stories, either statically (\textit{Annotation}) or dynamically (\textit{Linking}) and guide readers' attention to regions of visualizations that are central to the storytelling.
% Besides, the formats of common data stories can be automatically transformed into the ones with these two design methods~\cite{lai2020automatic}, which .

% To this end, we identify two most common techniques adopted in existing literature to connect visually separated text and visualization in data stories -- the most common format in journalism\cite{zhi2019linking, lai2020automatic}: explanatory annotation and interactive linking.

To evaluate the effectiveness of enhanced text-visualization integration, we designed and conducted a crowdsourcing study on Amazon Mechanical Turk with 222 participants to measure whether \textit{Linking} and \textit{Annotation} can affect people's awareness of misinformation and perceived credibility of text and visualization.
We carefully selected three data stories on popular topics that are flooded with misinformation as study materials.
For each story, we edited the original material to inject one piece of misinformation using one of the three methods identified in previous research: truncated axis (a case of deceptive visualization), inverted axis (another instance of deceptive visualization), and text-visualization contradiction (a kind of text-visualization misalignment)~\cite{pandey2015deceptive, kong2019trust}. 
Our experimental results showed that \textit{Linking} and \textit{Annotation} do enhance readers' awareness of misinformation and significantly lower the perceived credibility of the text or visualizations in the corresponding stories.
However, we also noticed most participants are still unaware of misinformation.
For participants who did not find misinformation in the given stories, we analyzed their subjective feedback on possible reasons and proposed design opportunities to combat misinformation.

% In summary, the key contributions of this paper are three-fold:
% \begin{itemize}
%     \item We explored how misinformation can be injected into a narrative visualization through a literature survey and summarized three causes of misinformation. 
%     \item We conducted a crowdsourcing study to test the possible effects of two defensive design methods (i.e., \textit{Linking} and \textit{Annotation}) on people's awareness of three types of misinformation in narrative visualization.
%     \item We performed a qualitative analysis of the participants' feedback and derived design implications for combating misinformation in data stories.
% \end{itemize}

\section{Related Work}
\subsection{Combating Misinformation}
% 关于misinformation的定义放在这里，以解释研究范围？还是放在sec 3?
Misinformation is commonly defined as any false or inaccurate information \cite{wu2019misinformation}.
Interpretation of ``information'' in this definition, however, may change in different scenarios, as it could mean facts provided or facts learned\cite{stevenson2010oxford}.
Kong et al. remove this ambiguity by thinking from the readers' perspective and define misinformation as ``information that leads to misperceptions about the facts''~\cite{kong2019trust}.
In this paper, we follow the definition by Kong et al. and consider incorrect or misleading information readers receive and consume.
% Researchers have tried to characterize such misinformation from aspects such as errors and bias\cite{bednar2008bias}. 
Despite the different research angles, there is a general consensus on the negative impact of misinformation~\cite{wu2019misinformation}.
% The prevalence of Internet and social media has made it very easy to circulate misinformation, making it a dire threat to the public in the digital ecosystem\cite{allcott2019trends}.
%Such definition of misinformation can have many sides such as disinformation, errors, bias and so on.
% combat misinformion之必要
%Regardless of which side of misinformation, it may have a negative impact on society. \color{red} citings...\color{black}
Therefore, methods to combat misinformation in the digital world are gaining increasing attention from the research community and the general public.
% \xm{[ref]}.

The most straightforward anti-misinformation approach is designing algorithms to automatically detect false or inaccurate facts disseminated online. 
Researcher have developed algorithms that utilize content data (e.g., text, image), context data (e.g., published time or location), and  propagation data (e.g., retweet and likes on social media) to detect misinformation\cite{wu2019misinformation}.
Still, there are concerns about the generality of current detection methods\cite{guo2019future}, and few existing algorithms take visualizations as input into their analysis.
% The awareness of misinformation have been also studied by psychologists for a long time.
% \cite{lewandowsky2012misinformation} ...

In the visualization area, researchers have studied misinformation in the forms of cognitive bias, visualization mirage, deceptive visualizations, etc. 
\cite{8585669,10.1145/3313831.3376420,pandey2015deceptive}.
Several works have experimented with means to combat such misinformation in ordinary visualizations.
For example, McNutt et al. \cite{10.1145/3313831.3376420} propose a metamorphic testing method to surface visualization mirage in visual analytics process.
% Wall et al. \cite{8585669} illustrate six metrics to detect cognitive bias based on users' interaction.
% Law et al. \cite{law2020causal} found that adding simple warnings to visualizations can reduce the credibility of misleading causal claims in QA systems.
% Hopkins et al\cite{hopkins2020visualint} present a technique that sketchily render erroneous visualization elements to help users identify chart construction errors.
Different from previous research, our work focuses on misinformation occurring in the production-consumption process of data stories and explore design alternatives to combat misinformation that come from information misalignment.

\subsection{Narrative Visualization}
% \todo{add refs}
% \review{
% R2: Kong et al, Extracting references between text and charts via crowdsourcing, CHI 2014
% }

Visualizations have been widely used to tell stories about data.
Segal and Heer\cite{segel2010narrative} introduced the term ``narrative visualization'' to cover such type of visualizations. 
Despite the varieties in genres, narrative visualizations generally combines (textual) narratives with interactive graphics\cite{segel2010narrative}.
%They also proposed a design space and concluded seven genres for narrative visualization.

Prior research around narrative visualization can be divided into three categories.
The first type of research focuses on exploring the design space of a particular genre or aspect of narrative visualization.
For example, Bach et. al ~\cite{bach2018design} conclude design patterns for data comics.
% Shu et. al\cite{shu2020makes} present a design space for data-GIFs and explore factors that make a data-GIF understandable.
% Brehmer et. al\cite{7581076} study the design space of timeline to tell stories in an event sequence.
% Stolper et al.\cite{stolper2016emerging} summarize visualization-driven storytelling techniques into high-level categories concerning communication and explanation, connection, navigation, and controlled exploration. 
The second type of research seeks to develop algorithms and systems to facilitate the authoring process of narrative visualizations.
Examples include Calliope~\cite{shi2020calliope} which automatically generates data stories from spreadsheets.
Kong et al.~\cite{kong2014extracting} also proposed a crowdsourcing-based method to create interactive articles.
The third type of research investigates factors that influence readers' experience to motivate better design of narrative visualizations.
For instance, McKenna et. al \cite{mckenna2017visual} look into how the reading flow of data stories affects readers' engagement.
% Liem et. al \cite{liem2020structure} compare the effects of structured and personal visual narratives on the attitude of readers.
Our work is close to the last type of research and aims to investigate factors that might influence readers' awareness of misinformation in data stories.

We are particularly interested in factors concerning narrative visualization design. 
Given the considerably large design space, Hullman and Diakopoulos~\cite{6064988} have proposed a framework to analyze what can affect users' interpretation of narrative visualization, which consists of four editorial layers, i.e. data, visual representation, textual annotation and interactivity. 
Among them, design techniques applied on the annotation and the interactivity layers can promote reader's digestion of the story content and assist in awareness building by enhancing the connection between text and visualizations within a data story~\cite{petty2012communication}.
% Among them, design techniques applied on the annotation and the interactivity layers can potentially help enhancing the connection between text and visualizations within a data story, can promote reader's digestion of the story content and assist in awareness building~\cite{petty2012communication}.
\revision{
For example, Kwon et al.~\cite{kwon2014visjockey} propose a technique that dynamically links text with relevant visualizations.
Such a technique can effectively guide readers through the author's intended narratives.
}
Zhi et al.~\cite{zhi2019linking} investigate two forms of text-visualization-integration methods -- layout and linking.
They found slideshow layout with linking can improve readers' recall of and engagement with a narrative visualization.
Wang et al.~\cite{wang2019comparing} compared different narrative visualization types and found increasing text-visualization integration can promote reader's understanding.
\revision{
Our study focuses on two design methods to connect textual and visualization components in data stories, i.e., \textit{Annotation} and \textit{Linking}.
Both of them are commonly used in designing magazine-style narrative visualization.
Moreover, there exist fast algorithms~\cite{lai2020automatic, kong2014extracting} to implement these two design methods, which make them easy to promote.
}

\section{Misinformation in Narrative Visualization}
\begin{table*}[h]
	\centering
	\caption{Possible causes of misinformation in every step of the production-consumption of narrative visualization. This table is not exhaustive but shows possible causes of misinformation. The lists in Analyzing and Visualizing steps are simplified from~\cite{10.1145/3313831.3376420}.}
	\label{tab:misinformation}
	{
		\begin{tabular}{lll}
			\toprule
			Step & Causes of Misinformation & Description \\
			\cmidrule[0.2pt]{1-3}
			Analyzing & Cherry Picking & Filter out data that is not conducive to the author's intended narrative\cite{cockburn2018hark} \\
		    \cmidrule[0.2pt]{2-3}
			& The Simpson's paradox & high-level aggregation of data leads to wrong conclusion\cite{guo2017you} \\
			
			\cmidrule[0.2pt]{1-3}
			Visualizing & Break Conventions & Create unusual charts that mislead people to analyze them with conventions\cite{correll2017black, pandey2015deceptive, lauer2020people} \\
		    \cmidrule[0.2pt]{2-3}
			& Concealing Uncertainty & Conceal the uncertainty in the chart to cover up the low quality of the data\cite{sacha2015role} \\
			
			\cmidrule[0.2pt]{1-3}
			Scripting &
			Text-visualization misalignment &
			The message of the text differs from that of the visualization it refers to\cite{kong2018frames, kong2019trust}\\
			\cmidrule[0.2pt]{2-3}
			& Text Wording &
			The degree of text intensity can bias readers' memory of graph\cite{newman2018effects} \\
			\cmidrule[0.2pt]{2-3}
			& Illusions of causality &
			The text makes incorrect causal inductions on chart information
			 \cite{matute2015illusions, law2020causal, matute2015illusions}\\
			 \cmidrule[0.2pt]{2-3}
			& Obmiting context &
			Omitting the context needed to understand the story
			\cite{doan2021misrepresenting} \\
% 			\cmidrule[0.2pt]{2-3}
% 			& Manipulating source &
% 			Misclaiming in the text that the data comes from an authority
% 			\cite{brennen2021beyond} \\
			\cmidrule[0.2pt]{2-3}
			& Manipulating order & 
			Manipulate the reading order through layout, resulting in order bias
			\cite{feenberg2017s} \\
			
			\cmidrule[0.2pt]{1-3}
			Arranging &
			Obfuscation &
			Make it difficult for readers to extract visual information through chaotic layout
			\cite{correll2017black} \\
			
			\cmidrule[0.2pt]{1-3}
			Reading &
			Personal bias & Political attitudes, beliefs and other personal factors lead to misperception of facts\cite{meirick2013motivated, garrett2014implications} \\
			\bottomrule
		\end{tabular}
	}
\end{table*}

McNutt et al. propose a pipeline of visual analytics to investigate how errors generated in each phase may undermine messages conveyed in the final visualization~\cite{10.1145/3313831.3376420}.
We are inspired to adopt a similar approach to obtaining a comprehensive view of the relationships between critical narrative visualization creation steps and the types of misinformation that can be introduced to mislead readers. % to dissect the data storytelling process and analyze where misinformation (following the definition in Sec 2.1) might come from in each stage.
To this end, we first construct a pipeline that dissects the production-consumption process of data storytelling (covered in subsection 3.1) and then conduct a literature survey to identify possible sources of misinformation (following the definition in subsection 2.1) at each stage of the pipeline (detailed in subsection 3.2). 

\subsection{Data Storytelling Process}

\begin{figure*}[tb]
 \centering
 \includegraphics[width=0.83\linewidth]{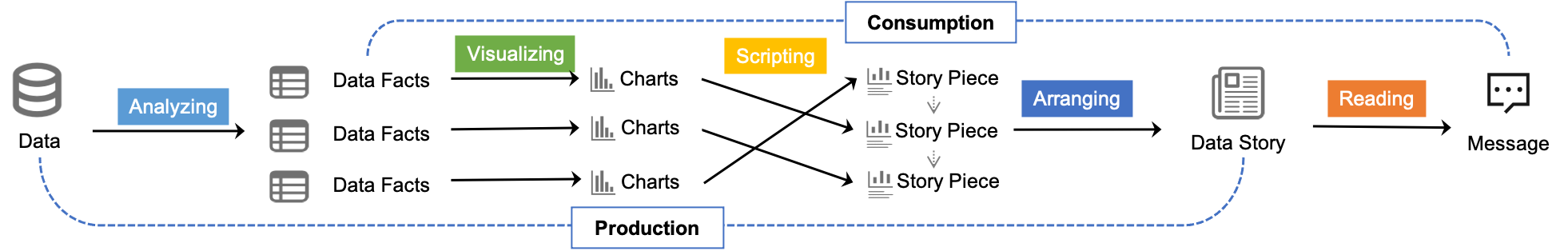}
 \caption{The production-consumption process of narrative visualization}
 \label{fig:process}
\end{figure*}

\revision{
Our pipeline is mainly adapted from Lee et al.'s visual storytelling process~\cite{7274435} while also motivated by previous works about visual analytics~\cite{10.1145/3313831.3376420, chen2018supporting}.
As we are more concerned with how possible falsified information injected in the production-consumption process and lead to readers' misperceptions of facts, we follow McNutt et al.'s~\cite{10.1145/3313831.3376420} practice and omit complex cycles and nested sub-process~\cite{7274435, chen2018supporting} to enable easy tracing of the origin of misinformation.
% Our proposed pipeline incorporates multiple previous works about visual analytics and visual storytelling~\cite{segel2010narrative, 10.1145/3313831.3376420, 7274435, chen2018supporting}.
% As we are more concerned with how falsified information injected in the production-consumption process may lead to readers' misperceptions of facts, we omit complex cycles and nested sub-process~\cite{7274435, chen2018supporting} to enable easy tracing of the misinformation.
}
% Our proposed pipeline incorporates McNutt et al.'s conceptual model of problems in the visual analytics process\cite{10.1145/3313831.3376420}, Lee et al.'s visual storytelling process\cite{7274435}, and Chen et al.'s framework of connecting visual analytics with storytelling\cite{chen2018supporting}.
% As we are more concerned with how possible errors or falsified information injected in the production-consumption process may lead to readers' misperceptions of facts, we follow McNutt et al.'s~\cite{10.1145/3313831.3376420} practice and omit complex cycles and nested sub-process~\cite{7274435, chen2018supporting} to enable easy tracing of the origin of misinformation.

Our pipeline describes the path from raw data to the perceived message by five main steps, namely, \textit{Analyzing}, \textit{Visualizing}, \textit{Scripting}, \textit{Arranging}, and \textit{Reading} (see Fig.~\ref{fig:process}). We further conceptualize the production process as steps involving authoring actions and the consumption process as steps of which the output will be directly perceived by readers.
According to this, \textit{Analyzing} is only related to the production process and \textit{Reading} is only associated with the consumption process.
\textit{Visualizing}, \textit{Scripting}, and \textit{Arranging} are involved in both the production and the consumption processes. The detailed definitions of these steps are as follows:

\textbf{\textit{Analyzing}} denotes the process that authors analyze the raw data to extract data facts that they intend to convey~\cite{chen2018supporting, 7274435}. In this step authors determine what to tell (and not to tell) in the final data story.
% which corresponds to the ``visual analytics'' part in Chen et al.'s\cite{chen2018supporting} framework and the ``Exploring Data'' phase in Lee et al.'s\cite{7274435} storytelling process.

\textbf{\textit{Visualizing}} characterizes the process that authors create charts to visually communicate the complied data facts to readers~\cite{10.1145/3313831.3376420}.
% After compiling a collection of data facts to present, , which we call ``Visualizing'' step as in McNutt et al.'s conceptual model\cite{10.1145/3313831.3376420}. 
Note that the charts here refer to the graphics readers actually see in \textit{the final data story}, not the ones that are used to analyze the raw data. 
% They are different from the exploratory plots authors draw in the Analyzing stage to get a better sense of data facts. Yet it is possible that the output charts of the Visualizing step are developed upon the draft plots used in the Analyzing step. %\xm{how does this step correspond to stages in other frameworks in the literature?}

\textbf{\textit{Scripting}} indicates the process that authors organize multiple data facts and their corresponding charts into a certain narrative structure, including \textit{ordering}, \textit{designing interaction}, and \textit{messaging}~\cite{segel2010narrative}.
The goal of this stage is to combine the scattered fragments of facts into a complete story to convey author-driven narratives~\cite{segel2010narrative,lee2021viral,6064988}.

% the composing process of textual contents for the data facts and their corresponding charts, which may include captions, annotations and a complete article~\cite{segel2010narrative,lee2021viral,6064988}.
% the process that authors organize multiple data facts and their corresponding charts into a certain narrative structure~\cite{segel2010narrative}. Textual contents are composed in this stage to tell author-driven narratives~\cite{segel2010narrative,lee2021viral,6064988}.
% To weave multiple data facts and their corresponding charts into a unified data story, authors would need to organize them in some sequence and provide associated textual content to tell author-intended narratives\cite{segel2010narrative,lee2021viral,6064988}.
% We refer to this step as ``Scripting'', and its goal is to combine the scattered fragments of facts into a complete story with a clear plot or storyline, making the whole piece easy to follow by readers\cite{hullman2013deeper, lee2021viral}.

\textbf{\textit{Arranging}} indicates the process that authors lay out multiple pieces of texts and/or charts in a scene and build the visual narrative flow for the readers~\cite{chen2018supporting, mckenna2017visual}.
Note that Arranging is concerned with the format of the data story but not the content.
% % TODO: phrasing
% After consolidating on a storyline in the Scripting step, authors would need to determine the specific positions of charts and text in the reading space.
% In other words, 
% We denote this step as ``Arranging''.
% Note that Arranging is only concerned with the story's format and does not make any edition to the content.

\textbf{\textit{Reading}} indicates the process that readers decoding information of the data story including the text and visualizations.
% by following the narrative of text and decoding visualizations that illuminate the narrative. 
The final message that readers derive is a result of the interplay between the presented information of the data story and their inherent knowledge, purpose and beliefs~\cite{canham2010effects, friel2001making}.

\subsection{Where does Misinformation Come from?}
% \todo{scientific rigor}
% \review{
% The paper presents the results of a quasi-survey of 17 papers and proposes three potential causes of misinformation. This subsection also feels largely unsubstantiated and lacks scientific rigor.
% }

% By surveying a collection of \xm{XXX} papers on \xm{XXX, XXX, XXX -- topical keywords} published in \xm{XXX, XXX, XXX -- venues e.g., TVCG, CHI, etc.}, 
\revision{
By searching for related publications in TVCG, CHI, and other relevant venues, we synthesize in detail 17 number of the most relevant literature. 
We identify possible misinformation causes in every step of the pipeline proposed above from the literature.
We present these causes and corresponding literature in Table.~\ref{tab:misinformation}.} Besides, we also consult works from journalism and psychology to survey more possibilities of causes of misinformation.
\revision{
Next, we put the resulting list of possible causes of misinformation in the context of narrative visualization production and consumption and divide them into three categories:
author-induced reader-undetectable misinformation (AIRUM), author-induced reader-detectable misinformation (AIRDM)  and reader-induced misinformation (RIM).
% and author-reader-co-induced misinformation\cite{bednar2008bias}.

\textbf{Author-Induced Reader-Undetectable Misinformation} 
By AIRUM, we mean false or inaccurate information injected into the production process through manipulations invisible to readers.
AIRUM mainly occurs in the Analyzing step.
For instance, authors might only pick those data that support their intention or even falsify data to meet their claims\cite{cockburn2018hark}.
Readers with external knowledge about the data story and its context might be able to find such AIRUM; however, it cannot be corrected based only on the information provided by the authors.

\textbf{Author-Induced Reader-Detectable Misinformation}
We define AIRDM as miscommunication rooted in the production process but can be detected by careful readers.
Such misinformation might concern Visualizing, Scripting, and Arranging steps as the intersection of the production and consumption processes illustrated in Fig.~\ref{fig:process}.
Authors' decisions in these steps are directly reflected in the output, and skeptical readers can spot the misleading elements.
Deceptive visualization is a well-known cause of AIRDM in the Visualizing step.
The author may truncate the axis or take area as the quantity that misleads readers intentionally or unintentionally~\cite{pandey2015deceptive}.
Readers can capture the existence of such misinformation by comparing the difference(s) between the deceptive chart and the conventional ones.
Text-visualization misalignment is a typical example of AIRDM that occurs in the Scripting step~\cite{kong2019trust}.
The author may manipulate the text-chart correspondence to (mis)guide the reader's attention, but a vigilant reader can find misalignment through close inspection.

\textbf{Reader-Induced Misinformation}
RIM refers to the misperception that happens during the consumption process and depends on how the visual argument(s) are read in real-world situations.
This kind of misinformation relates primarily to the Reading step.
The value, belief, and the information exposure environment of an individual could all cause misreadings~\cite{meirick2013motivated, garrett2014implications}.
RIM impact readers differently.
Authors' decisions in the production process can change the tendency of a message being more deceptive or not, but cannot fully determine whether the story will mislead readers~\cite{roozenbeek2020susceptibility}.
% its actual effect on misleading readers as well as the reliability of that effect\cite{roozenbeek2020susceptibility}.
% Deceptive visualization is a well-known cause of author-reader-co-induced misinformation in the Visualizing step.
% The author may create a deceptive visualization by truncating or reversing the axis intentionally or unintentionally~\cite{pandey2015deceptive}.
% Readers can capture the existence of such misinformation by comparing the difference(s) between the deceptive chart and the conventional ones.
% Author-reader-co-induced misinformation may occur in the Scripting step as well, and one typical example is text-visualization misalignment~\cite{kong2019trust}.
% The author may manipulate the text-chart correspondence to (mis)guide the reader's attention, but a vigilant reader can find misalignment through close inspection.

% As in the Arranging step, obfuscation\cite{correll2017black} can be employed by authors to mislead readers, placing visualizations that are not conducive to their narratives in an inconspicuous position. This may be detected if one (deliberately) search for the message.

Combating AIRUM requires a more rigorous authoring method (e.g., metamorphic testing\cite{10.1145/3313831.3376420}).
To fight RIM, promoting education on data and visualization is one of the directions currently under consideration by academics\cite{ecker2014correcting}.
As for mitigating AIRDM, a plausible approach is presenting the data story in a way that allows readers to scrutinize it easier\cite{roozenbeek2020susceptibility}. 
Therefore, in the scope of this paper, we focus on AIRDM and conduct a crowdsourcing study to explore whether different narrative visualization designs may affect readers’ awareness of such misinformation.}

\section{Study Design}
% We design a crowdsourcing user study to explore the possible effect of enhanced text-visualization integration on readers' awareness of misinformation in data stories.

We designed a crowdsourcing user study to explore the possible effects of two widely-used design methods in narrative visualizations, explanatory annotation (\textit{Annotation}) and interactive linking (\textit{Linking}), on readers' awareness of misinformation in data stories.
The misinformation we tested in the study is the type of author-induced reader-detectable misinformation.
More specifically, two types of deceptive visualization~\cite{pandey2015deceptive} and one type of text-visualization misalignment~\cite{kong2019trust} were used.

% Relying on one-sided information to understand data stories is an essential reason for readers unaware of misinformation.
% For example, the study by Kong et al.~\cite{kong2019trust} showed that a majority of participants remembered false information in the title rather than the accurate information in the visualization.
% Lauer and O'Brien~\cite{lauer2020people} found that people over-rely on the shape of the line in line charts even though some understand the line may exaggerate the message.
% To counter such a problem, we hypothesize that users can find the misinformation in the narratives more efficiently by balancing attention to different components of a data story.
% To this end, the two design methods that were used in the study(\textit{Annotation} and \textit{Linking}) bridge visualizations and text, two critical components for storytelling, visually and guide users' attention between regions of text and visualization that are central to the storytelling.
% The functions of guiding attention and the successful usages of \textit{Annotation} and \textit{Linking} in narrative visualization~\cite{segel2010narrative, zhi2019linking} motivate us to test them as the candidates of defensive design methods to misinformation.

\subsection{Story Selection}

We collected candidate data stories for the study either from well-known media (e.g., New York Times)
% \footnote{\url{https://www.nytimes.com/interactive/2020/world/americas/brazil-coronavirus-cases.html}}
% \footnote{\url{https://www.nytimes.com/interactive/2017/06/01/climate/us-biggest-carbon-polluter-in-history-will-it-walk-away-from-the-paris-climate-deal.html}} 
or from influential information publisher (e.g., Our World in Data).
% \footnote{\url{https://ourworldindata.org/obesity}}). 
\revision{
In the selection process we took several factors into careful consideration.
Most importantly, we focused on topics like health and climate changes where misinformation prevail~\cite{roozenbeek2020susceptibility, chou2018addressing, cook2018deconstructing}.
Then, we chose interesting and relevant stories to motivate reading.
}
Last but not least, we require that the text in a given story must reference some part(s) of the visualization to increase the likelihood that readers would attend to both types of contents.
We also need the original data behind the stories to ensure the feasibility of embedding misinformation into the selected narrative visualization. 

After several rounds of careful screening and discussions, we finally selected three short stories from three articles for our study:
\begin{itemize}[noitemsep,topsep=0pt,parsep=0pt,partopsep=0pt]
    \item \textbf{COVID-19 story}\footnote{\url{https://www.nytimes.com/interactive/2020/world/americas/brazil-coronavirus-cases.html}}: it presents the changes of infected cases during the COVID-19 epidemic in Brazil from June 2020 to January 2021 using a \underline{line chart}. The text refers to the visualization to explain public policies, like reopening tourist attractions, adopted by the Brazil government. 
    \item \textbf{Obesity story}\footnote{\url{https://ourworldindata.org/obesity}}: it stresses that obesity has become a critical risk factor for increasing mortality, by comparing the significance of obesity as a cause of death to other risk factors like smoking in a \underline{bar chart}.
    \item \textbf{Carbon Emission story}\footnote{\url{https://www.nytimes.com/interactive/2017/06/01/climate/us-biggest-carbon-polluter-in-history-will-it-walk-away-from-the-paris-climate-deal.html}} : it argues that developing countries shouldn't rely on fossil fuels for development as what the developed countries have done in the past. A \underline{bar chart} is used to compare the per person carbon emissions (metric tons CO2) of several countries in 2014. China is highlighted to illustrate the opinion.
\end{itemize}

We did not use the entire article but extracted sections that related to the visualizations.
We also slightly edited the sections to make the three stories have similar length (COVID-19 \& Obesity: 117 words; Carbon Emission: 109 words).

\subsection{Visualization Design}
\label{vis-design}

\begin{figure*}[!htb]
  \centering
  \includegraphics[width=0.93\linewidth]{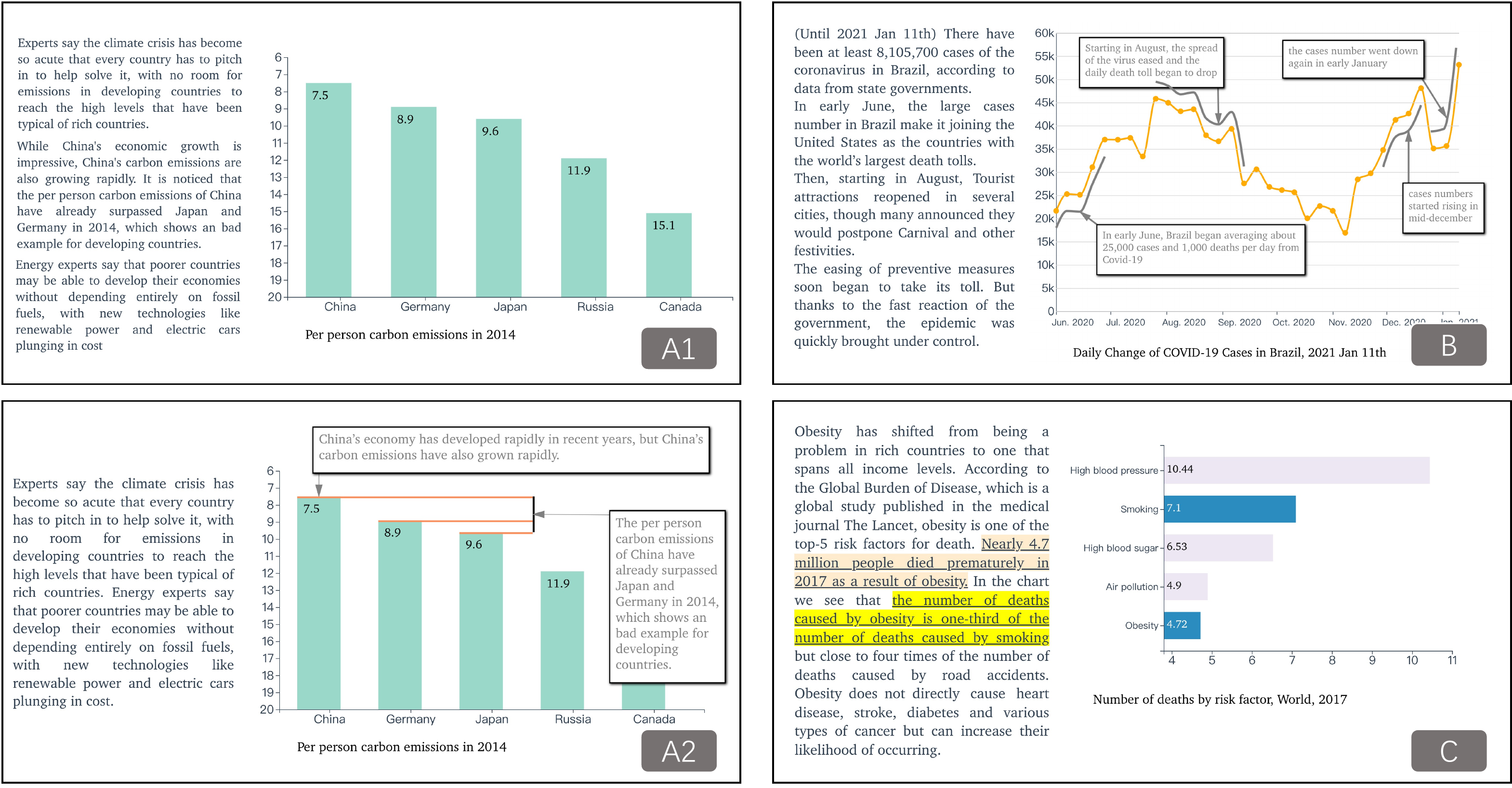}
    \caption{\revision{
    (A1) the Carbon Emission story under the \textit{Static} condition; (A2) the Carbon Emission story under the \textit{Annotation} condition; (B) the COVID-19 story under the \textit{Annotation} condition; (C) the Obesity story under the \textit{Linking} condition. 
    }
    % The columns from left to right show the Carbon Emission story, the COVID-19 story and the Obesity story. A1, A2, A3 show the Static condition. B1, B2, B3 show the Annotation condition. C1, C2, C3 show the Linking condition.
    }
  \label{fig:Interface}
\end{figure*}

For each selected story, we obtained its data from the source website. 
\revision{
We then reproduced the narrative visualizations. 
As the three stories' original text formats and layouts differ greatly, we unified their styles in our reproduced version by referencing online article styles in New York Times.
For visualizations, we only included elements concerned with the text we used in the study to avoid noise.
}
This process resulted in a line chart for the COVID-19 story, a horizontal bar chart for the Obesity story, and a vertical bar chart for the Carbon Emission story as in Fig.\ref{fig:Interface}.
Then, building on this basic data visualization, we designed and implemented the three conditions for our experiment.

\begin{enumerate}[noitemsep,topsep=0pt,parsep=0pt,partopsep=0pt]
    \item Static Illustrated Text (\textit{Static}). Text paragraphs are placed on the left side of the page, and the visualization is displayed on the right. Text and the visualization are separated, and for all stories, they take similar space in the experimental website.
    \item Explanatory Annotation 
    (\textit{Annotation}). Text which has no connection to the visualization is put on the left. Each text relevant to a particular part of the visualization is formatted as an annotation enclosed in a textbox. Every annotation locates near the corresponding part in the visualization and has an arrow pointing to it. The layout of annotations and arrows is crafted manually to avoid blocking essential contents of visualization. No hints are implying the reading sequence of annotations.
    \item Interactive Linking (\textit{Linking}). The layout of text and visualizations is identical to that in the illustrated text condition. Sentences that reference the visualization are highlighted. When users hover over the highlighted text, its corresponding part of the visualization will be spotlighted by changing to colour more salient than that of the rest.  %The color of the part that text refer to is deeper and those text not refer to are lighter. 
\end{enumerate}

\subsection{Misinformation Injection}

We concentrate on AIRDM in the scope of this work.
For each story in our study, we select one type of misinformation to inject into its production-consumption pipeline as described in the previous section.
In particular, we choose to introduce misinformation caused by deceptive visualization and text-visualization misalignment as they have received wide discussions in the research community and have already demonstrated negative impact on the society \cite{pandey2015deceptive, kong2019trust, lauer2020people}. 

To be more specific, we instill \textbf{Contradictory Slant} -- a form of text-visualization misalignment -- into the COVID-19 story by adding a sentence stating that the number of infections is declining attributed to the Brazilian government while the number shown in the line chart is actually rising.  %to its line chart.
The contradictory slant is an easy-to-spot text-visualization misalignment; still, many people fail to find such misinformation~\cite{kong2019trust}.
% According to a prior study done by Kong et al.\cite{kong2019trust}, the contradictory slant is an easy-to-spot text-visualization misalignment; still, most participants ($68-80\%$) fail to recognize such misinformation.
%We inject the contradictory slant into the COVID-19 story by adding a sentence stating that the number of infections is declining and attribute it to the Brazilian government when the number of cases shown in the chart is rising in January 2021. 

We apply two forms of deceptive visualization to the Obesity story and the Carbon Emission story, respectively, namely \textbf{Truncated Axis} and \textbf{Inverted Axis}.
Both of them are popular deceptive techniques for visualization and have been thoroughly discussed in previous works\cite{pandey2015deceptive}.
In the Obesity story, by truncating the X-axis of the horizontal bar chart, we make the risk of obesity seem much smaller (in comparison with the real number) than that of smoking to go with the claim in the text.
In the Carbon Emission story, we make China's per capital carbon emissions appear the highest in the bar chart while its value is actually the smallest by inverted the Y-axis. We compared China with Japan and Germany in their carbon emission control performance in the text.
\revision{
Note that in narrative visualization, \textbf{Truncated Axis} and \textbf{Inverted Axis} can be considered as special forms of \textbf{Contradictory Slant}, as the text in all three types of misinformation is not aligned with the visualization.
However, the deceptive visual elements in \textbf{Truncated Axis} and \textbf{Inverted Axis} make it more challenging for readers to find the misinformation.
}

% \begin{table}[!htb]
% 	\centering
% 	\caption{Injected Misinformation Types}
% 	\label{tab:mistypes}
% 	{
% 		\begin{tabular}{@{}llll@{}}
% 			\toprule
% 			& Contrad. Slant & Truncated Axis & Inverted Axis \\ \midrule
% 			vis & honest & deceptive & deceptive \\
% 			text-vis & misaligned & aligned & aligned \\
% 			text & incorrect & incorrect & incorrect \\
% 			\bottomrule
% 		\end{tabular}
% 	}
% \end{table}

% \revision{
% The similarities and the differences between the three types of misinformation can be best summaried in Table.~\ref{tab:mistypes}.
% }

As mentioned above, the text component of each data story contains a single sentence that explicitly states the misinformation.
\revision{Although the misleading sentence is only part of a complete data story, it is central to the conveyed message, either to convince readers that the epidemic is under controlled (COVID-19 story), or to lower the dangerous of obesity (Obesity story), or to exaggerate China's negative impact (Carbon Emission story).}

\revision{
Besides, we assign each misinformation to a specific story to avoid finding it too easy. 
For example, suppose we inject \textbf{Inverted Axis} to the Obesity story and rank the number of death caused by air pollution higher than that by high blood pressure. In that case, readers can find the misinformation easily by common knowledge.
We want to observe the impact of the design methods used and lower the impact of other factors.
}

\begin{figure*}[h]
  \centering
  \includegraphics[width=0.75\linewidth]{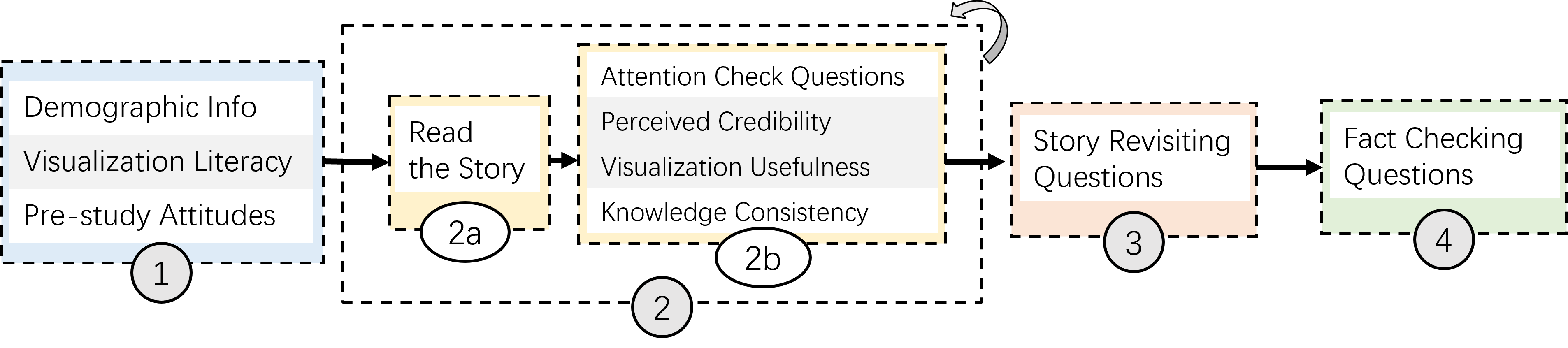}
    \caption{The procedure of the study. Note that each participant went through stage \textbf{2} three times as they read three stories in total.}
  \label{fig:procedure}
\end{figure*}

\subsection{Participants}
We conducted the between-subjects experiment on Amazon Mechanical Turk (MTurk), a popular crowdsourcing platform\cite{borgo2018information} with lay public.
%We choose MT as our experiment platform because...
We set the following inclusion criteria for selecting participants to ensure the quality of the study: 1) number of Human Intelligence Tasks (HIT)' approved greater than 1000; 2) 97\% or above HIT approval rate, and 3) self-reported to be in the United States. 
In total we received the responses from 313 unique participants.
Based on the filtering criterion stated in Sec. \ref{sec:measure}, we filtered out 91 responses.
% The filtering rate is 28.12\%, which is closed to the one in previous studies that were also conducted on MTurk~\cite{}.
The resulting dataset includes 222 responses, with 98 females and 124 males.
Each condition was tested by 74 participants.
All participants signed a consent form that present the procedure of the study and data we would collect prior to the study.
The mean age of participants was 40 years old (20 to 65).
Majority of the participants (72.072\%) have a bachelor's degree or above.
Each HIT took approximately 20 minutes to complete, and the participants received a compensation of US \$2.5.

\subsection{Procedure}
\label{procedure}
Participants went through our study following the steps as presented in Fig. \ref{fig:procedure}.
Each of them read the three data stories under one random-assigned condition (i.e., \textit{Static}, \textit{Linking} or \textit{Annotation}).
They completed the tasks by accessing our study website that was posted on MTurk.
The participants were provided with an online consent form after the study website is loaded.
After they agreed on the terms, we asked questions regarding the demographic information, \revision{ self-reported visualization literacy level}, and the attitudes towards the three topics covered in the stories~(as shown in step 1 in Fig. \ref{fig:procedure}).
Then, they proceeded to complete the tasks in the following phases:

\textbf{Story reading phase} As shown in step 2 in Fig.\ref{fig:procedure}, the participants were required to read the three data stories and answer the follow-up questions after reading each story.
The order of stories and the assignment of conditions were counter-balanced among participants.
The participants could spend as much time on each story but no shorter than 30 seconds (mandatory minimum viewing time). 
Note that after reading each story, we did not ask participants to report detected misinformation because we didn't want participants to look for misinformation on purpose in subsequent stories.
Such questions were asked in the later phases.

\textbf{Story revisiting phase} After participants had read all stories and answered corresponding questions, we presented the three stories again and asked them to report if they had found any misinformation (denoted as Story Revisit questions).
If participants claim to have detected misinformation, we asked them to specify the sentence(s) that contain the identified misinformation (step 3 in Fig. \ref{fig:procedure}). 
In the instructions of this phase, we emphasized no need to re-read the stories to complete these questions.

\textbf{Fact checking phase} After revisiting all three stories, participants were introduced to the facts and the methods to manipulate them (denoted as Fact-Checking questions).
They then were asked to reflect on whether such misinformation is easy to spot (step 4 in Fig. \ref{fig:procedure}).
If they do not find the misinformation during their previous reading, we invited them to write down the reasons.

\subsection{Measures}
\label{sec:measure}
We collected four groups of data from the online study:
\begin{itemize}[noitemsep,topsep=0pt,parsep=0pt,partopsep=0pt]
    \item pre-study questionnaire;
    % \item mouse moving logs data that was recorded when participants read stories, including the coordinates of the mouse pointer every 3 seconds.
    \item participant perceptions to the reading experience of each story;
    \item participant-claimed detected misinformation;
    \item qualitative feedback after the fact-checking phase.
\end{itemize}
We measured the participant perceptions to the reading experience via questions covering the four aspects: 
(1) attention checking. 
As suggested by Borgo et al.~\cite{borgo2018information}, it is important to check if the crowd workers pay attention to the tasks.
Hence, for each story, we asked two multiple-choice recall questions and an open-ended story-summairzation question to test if the participant read the story carefully. We only accepted submissions from participants who answer at least \textit{three} \revision{(out of six from all three stories)} multiple-choice questions correctly and provide a informative story summary with no fewer than five words;
(2) visualization usefulness. We asked participants to self-evaluate the usefulness of the visualization of the story they just read via a five-points likert scale.
(3) perceived credibility.
We adapted the credibility scales from Kong et al.~\cite{kong2019trust} to measure the perceived credibility to the text and visualization.
More specifically, the participants rated the text and the visualization of the just-read story from five perspective:  accuracy, fairness, trustworthiness, bias, and completeness. 
We toke average of the rates as the perceived credibility in the subsequent analysis.
(4) knowledge consistency. The prior knowledge and belief can influence the reading experience of pariticipants~\cite{johnston1984prior, kong2019trust}. Thus, we asked participants the level of consistency between the content of the story and their prior knowledge in seven-points likert scale to analyze the influence of prior knowledge to the results.

\section{Results}

\subsection{Awareness of Misinformation}
\label{sec:results-awareness}

\begin{figure*}[h]
  \centering
  \includegraphics[width=0.95\linewidth]{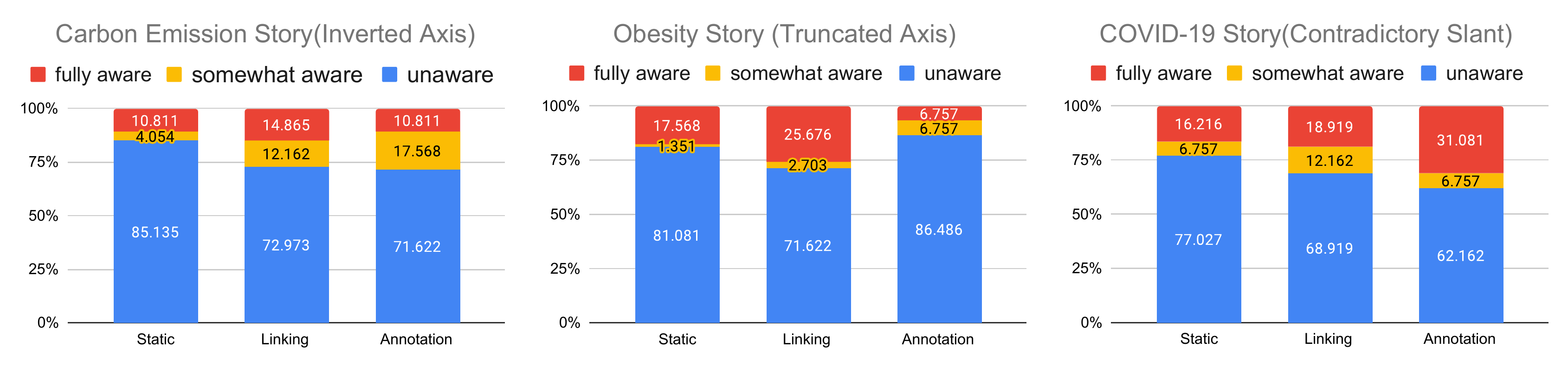}
    \caption{The percentage of participants with different degrees of awareness in each story under the \textit{Static}, \textit{Annotation} and \textit{Linking} conditions}
  \label{fig:Awareness}
\end{figure*}

% In the story revisiting phase, We asked the participants if they had detected any misinformation.
% By looking into the data, we found that although some participants stated that they did not find misinformation in this phase, however, they claimed in the subsequent fact-checking phase that they did find the injected misinformation, but as they misunderstood our questions or for other reasons
% For example, 
% \begin{displayquote}
% I did see that inconsistency between the text and the visualization, but did not compute that it was an error, which was wrong of me.

% I noticed this in the first reading of the article but by the time you asked about misinformation I started second guessing myself and thought perhaps I wasn't understanding the chart correctly.
% \end{displayquote}

To investigate whether participants were aware of the misinformation in the given data stories,
we coded participants' responses to the questions in the story revisiting and fact-checking phases (see subsection 4.5) and labeled their level of awareness based on the following criteria:
\begin{enumerate}[noitemsep,topsep=0pt,parsep=0pt,partopsep=0pt]
    \item Participants who stated finding the misinformation in the story revisiting phase and also gave the right sentence that contains that misinformation were labelled as ``\textbf{fully aware}".
    \item Some participants did not imply in the story revisiting phase that they found misinformation but in the fact-checking phase reported to have noticed the misinformation.
    For example, ``\textit{I did see that inconsistency between the text and the visualization, but did not compute that it was an error, which was wrong of me}''.
    We labeled such participants as ``\textbf{somewhat aware}'', if there exists evidence indicating that they hold reasonable suspicion.
    % while reading after checking their answers to other questions.
    Such evidence includes but is not limited to doubting (certain part of) the chart in their replies to the story-summarization questions, giving low credibility ratings ($\leq 4$) in the perceived credibility questions, etc.
    \item For any cases other than 1 and 2, we label them as ``\textbf{unaware}''.
\end{enumerate}

One author and one PhD student with psychology background coded the data.
After coding the first 100 responses, they met to compare the labeling results and resolve any conflicts through discussion.
Then they coded the rest data.
An inter-rater reliability analysis using the Kappa statistic shows high reliability ($\kappa = 0.84$).

% \subsubsection{Effects of \textit{Annotation} and \textit{Linking}}
Fig.~\ref{fig:Awareness} shows the distribution of participants' awareness levels under different conditions in each story with a specific form of misinformation injected.
The majority of participants ($62.162-86.486\%$) in all conditions failed to detect the misinformation.
\revision{
Despite that, we found a significant marginal impact of the conditions on participants' awareness of misinformation ($\chi^2(2) = 5.53$, p=0.063) across the three misinformation types using the Kruskal-Wallis test.}
We further employed Dunn's post-hoc tests with Holm correction for pairwise comparisons.
\revision{The results show that participants' awareness of misinformation under the \textit{Linking} condition is significantly higher than that under the \textit{Static} condition (Z = -2.29, p = 0.033).
But we did not find significant effects of \textit{Annotation} over the awareness.
}

Considering the three stories with different forms of misinformation separately, we compared the awareness under the \textit{Annotation} and \textit{Linking} conditions with the one under the \textit{Static} condition by running Mann-Whitney's U test.
We found that in COVID-19 story with contradictory slant, \textit{Annotation} has a significant effect on participants' awareness of misinformation (U=2303.5, p=0.039, r=0.159).
No other statistical significance was found.

The above results show that although the effects are limited, \textit{Annotation} and \textit{Linking} did have promotions to participants' awareness of misinformation.
\textit{Linking} is generally more effective than \textit{Annotation} in improving participants' awareness.
We can see from Fig. \ref{fig:Awareness} that in all three types of misinformation, the number of participants who were fully aware of misinformation under \textit{Linking} is higher than that under \textit{Static}. 
But for \textit{Annotation}, although it had an advance in the COVID-19 story with contradictory slant, in Obesity story, it is even worse than \textit{Static}.

\subsection{The Perceived Credibility to Text and Visualization}

After the participants read each data story, they scored their perceived credibility to the text and visualization.
In this section, we present our findings deriving from analysing participants feedback.

\begin{table*}[h]
	\centering
	\caption{ANOVA tests and post hoc comparisons for the three conditions under the different types of misinformation. The post hoc comparisons are conducted only when we found significance in the ANOVA tests.}
	\label{tab:postHocComparisons-Interaction}
	{
		\begin{tabular}{llll|lllll}
			\toprule
			Story Type & Misinfo. Type & ANOVA-p$_{text}$ & ANOVA-p$_{vis}$ & Comparison & $\Delta_{text}$ & post hoc-p$_{text}$ & $\Delta_{vis}$ & post hoc-p$_{vis}$  \\
			\cmidrule[0.4pt]{1-9}
			COVID-19 & \textit{Contrad. Slant} & 0.006** & 0.310 & \textit{Static} - \textit{Annot.} & 0.646 & 0.004** & - & - \\
			   & & & & \textit{Static} - \textit{Linking} & 0.405 & 0.111 & - & - \\
			   & & & & \textit{Annot.} - \textit{Linking} & -0.241 & 0.457  & - & - \\
			\cmidrule[0.4pt]{1-9}
			Carbon & \textit{Inverted Axis} & 0.023* & 0.059* & \textit{Static} - \textit{Annot.} & 0.435 & 0.041* & 0.503 & 0.065* \\
			   & & & & \textit{Static} - \textit{Linking} & 0.419 & 0.051* & 0.403 & 0.167 \\
			   & & & & \textit{Annot.} - \textit{Linking} & -0.125 & 0.721 & -0.100 & 0.894 \\
			\cmidrule[0.4pt]{1-9}
			Obesity & \textit{Truncated Axis} & 0.341 & 0.586 & - & - & - & - & - \\
			\bottomrule
			% \addlinespace[1ex]
			% \multicolumn{6}{p{0.5\linewidth}}{\textit{Note.} P-value adjusted for comparing a family of 3} \\
		\end{tabular}
	}
\end{table*}

\subsubsection{Does the perceived credibility differ?}
We perform MANOVA tests to investigate the impact of condition (\textit{Static}, \textit{Linking}, and \textit{Annotation}) on the two sets of credibility scores (for text and visualization respectively as mentioned in subsection \ref{sec:measure}, also see Table. \ref{tab:credibility}).
Across different forms of misinformation, we could see that the condition has a significant effect on the credibility scores (Approx. F=4.658, Wilks' $\lambda$=0.972, $p$=0.001). 
ANOVAs revealed that the condition has significant effect on the text and visualization credibility scores (p$\leq$0.001 and p=0.022, correspondingly).
\revision{
A post hoc Tukey test showed that the text credibility scores in the \textit{Annotation} and \textit{Linking} conditions are significantly \textbf{lower} than that in the \textit{Static} condition ($p \leq $0.001, d = -0.367; $p = $0.024, d = -0.321); the visualization credibility scores in the \textit{Annotation} condition is significantly \textbf{lower} than that in the \textit{Static} condition ($p$ = 0.016, d = -0.261).
No other significant differences were found.
}
% The follow-up Tukey's posthoc tests suggested that the text credibility scores in the \textit{Annotation} condition (M=4.77, SD=1.16) and in the \textit{Linking} condition (M=4.89, SD=1.01) are significantly \textbf{lower} than that in the \textit{Static} condition (Mean=5.14, SD=1.06) ($p\leq $0.001 and $p=$0.024, correspondingly).
% The visualization credibility score in the \textit{Annotation} condition (Mean=4.78, SD=1.26) is significantly \textbf{lower} than that in the \textit{Static} condition (Mean=5.08, SD=1.15) ($p$=0.016).
% No significant effects were found between the \textit{Linking} condition (Mean=4.88, SD=1.10) and the \textit{Static} condition regarding the visualization credibility score, and between the \textit{Annotation} condition and the \textit{Linking} condition regarding both credibility score.

\begin{table}[!htb]
	\centering
	\caption{Means and standard deviations of the credibility scores}
	\label{tab:credibility}
	{
		\begin{tabular}{@{}lll@{}}
			\toprule
			Condition & Text Credibility & Vis Credibility \\ \midrule
			Static & M = 5.14, SD = 1.06 & M = 5.08, SD = 1.15 \\
			Annotation & M = 4.77, SD = 1.16 & M = 4.78, SD = 1.26 \\
			Linking & M = 4.89, SD = 1.01 & M = 4.88, SD = 1.10 \\
			\bottomrule
		\end{tabular}
	}
\end{table}

We further investigated the effects of the conditions under the three types of misinformation separately.
The results are shown in Table \ref{tab:postHocComparisons-Interaction}.
We found that \textit{Annotation} significantly differs to \textit{Static} on the text credibility scores of the stories with \textit{Contradictory Slant} and \textit{Inverted Axis}, and on the chart credibility score of the story with \textit{Inverted Axis}.
Besides, with \textit{Inverted Axis}, \textit{Linking} also had a marginal significant effect on the text credibility compared to \textit{Static}.
\revision{
The above results show that \textit{Annotation} and \textit{Linking} can lower the perceived credibility of the data stories that contain misinformation.}

Worth mentioning, under all three conditions, the average text and visualization credibility scores across different misinformation lean to be positive (above 4).
This result is consistent with Sec. 5.2.1, where most participants did not report any (valid) misinformation.

\begin{figure}[!h]
  \centering
  \includegraphics[width=0.7\linewidth]{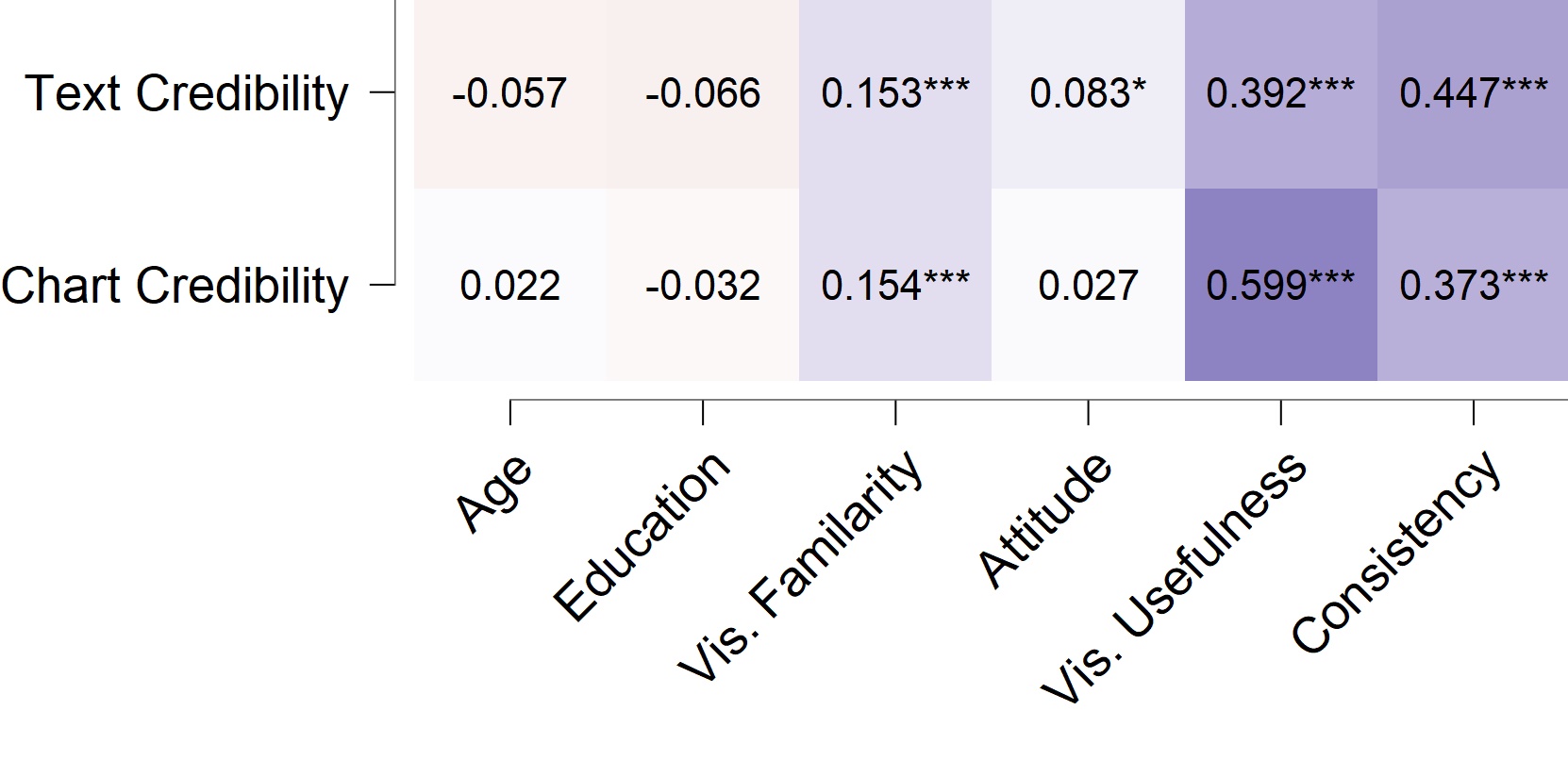}
    \caption{Correlations between six factors and two credibility scores}
  \label{fig:heatmap}
\end{figure}

\subsubsection{Compare the two types of perceived credibility}
We compared the credibility scores of text and visualization using Wilcoxon signed-rank tests for each type of misinformation.
Kong et al.~\cite{kong2019trust} found that for visualizations with title-visualization misalignment, people consider the visualization to be more credible than the text.
\revision{
For \textit{contradictory slant}, which was also used by Kong et al.~\cite{kong2019trust}, our results repeat their findings:
The text credibility (Mdn = 5.00) is significantly lower than the visualization credibility (Mdn = 5.20), Z = -3.64, p $\leq$ 0.001. 
But for stories with deceptive visualizations, we did not find similar significant differences.
}
% However, with deceptive visualization, the mean of the text credibility is slightly higher than the visualization credibility, and in particular, there is a marginal significance under the \textit{Inverted Axis}.
% Note that in our study, the text of the data story with deceptive visualization also contains misinformation, which refers to the misleading parts in the visualization.
% Our results imply that people can trust text more in a data story containing misinformation in both text and visualizations.

% We further look into whether the perceived text and chart credibility are different.
% % Fig.\ref{fig:credibility} shows the distributions of text and chart credibility scores for the three stories.
% As a Shapiro–Wilk test suggests a deviation of the data from normality, we conducted Wilcoxon signed-rank tests for each story.
% The results indicate that the text credibility is significantly higher than the chart credibility in the Carbon Emission story ($p=0.02$),
% In the Obesity story, there is a similar trend, but the significance is only marginal ($p=0.054$).
% In the COVID-19 story, in contrast, we observe a reversed pattern with the chart credibility significantly higher than the text credibility ($p<0.001$).

\subsubsection{Does the perceived credibility correlate to other factors?}
We are also interested in whether other factors like age, education, visualization literacy, attitudes, visualization usefulness and knowledge consistency will affect the perceived credibility.
Hence, we computed Spearman's rank correlations between these factors and the two credibility scores.
The results have been shown in Fig. \ref{fig:heatmap}.
We found little correlation between age, education,  self-reported visualization literacy (``Vis. Familarity'' in Fig.\ref{fig:heatmap}), attitude and the credibility scores.
These results are partially consistent with previous research~\cite{pandey2015deceptive, lauer2020people}.
The exception is attitude, which was found having effects on perceived bias by Kong et al.~\cite{kong2019trust}.
We believe this may result from the differences of stories used the two studies, which is worthy future study. 

We found a positive correlation between the visualization usefulness and the two credibility scores, especially the visualization credibility score (p$\leq$0.001).
Understandably, the participants who thought the text and visualizations were less credible rated the visualization as less useful.
However, it is surprising that the participants who found the visualization useful cannot find it less credible.
A possible reason is that the power of the misleading text induces the participants to ignore the errors in the visualization.
Besides, there is a weak positive correlation between the knowledge consistency and the credibility scores, consisting with previous research~\cite{johnston1984prior}.

\subsection{Why are Readers Misled?}

For participants who do not discover misinformation, we collected their subjective feedback on possible reasons and grouped them into seven categories using thematic analysis~\cite{braun2012thematic} (see Table~\ref{tab:reason}).

The most frequently mentioned reason is that their attention is not on elements related to misinformation (45.21\%).
For example, in the Obesity story with truncated axis, some participants ignored the numbers tagged on each bar: 
``\textit{I had not looked at the values for each of the bars at the initial viewing, as I was primarily focusing on the heights of the bars and comparing it to one another}''.
Also, in the COVID-19 story, some participants paid less attention to information near the end of the story: 
``\textit{It was not easy [to find the misinformation] because it was at the end of the paragraph. My mind was not fully focused on the information knowing that it was like a closing sentence}''.

Participants also identified text as an important factor that affects their understanding of the data story (``Excessive trust in text'' in Table. \ref{tab:reason} ).
This finding is consistent with the claim of Kong et al.~\cite{kong2019trust} that text has a strong power in manipulating the receiving information of data stories.
Even though some participants sensed something might go wrong with the chart, they did not consider it as misinformation: 
``\textit{I saw that the text and chart didn't match, but I assumed the text was correct and the chart had been printed wrong}''.

Misinterpretation of the visualizations is another reason.
For example, some participants failed to decode one dimension of the data points:
``\textit{The problem here is seeing where the line points match the date along the x-axis. It is not easy to see how the dates align with the line drawn}''.
As stated in Section 4.2, the visualizations used in the study are closed to what has been used on public online news.
Hence, this reason shows the importance of media improving the comprehensibility of visualizations in their publishing stories.

% The reason ranked second of the COVID-19 story with contradictory slant is 
% misinterpretation of the line chart, e.g., ``The problem here is seeing where the line points match the date along the x-axis. It is not easy to see how the dates align with the line drawn.''

Participants also reported Other reasons, include lacking prior knowledge (e.g, ``\textit{I think I was fooled by} [my prior impression of] \textit{Canada and somehow thinking of it as being full of empty space and clean air compared to the others}''), over-reliance on conventions of visualization (e.g, ``\textit{I don't know how one would be able to know that the y-axis is inverted without being told so}''), misunderstanding the reference relationship (e.g, ``\textit{I felt like the text was focused on the early January reaction to the December spike}''), and lack of intention to find the misinformation (e.g, ``\textit{This is not easy to spot and I was not really looking for misinformation}'').
We discuss the design implications from these reasons in Section 6.

\begin{table}[h]
	\centering
	\caption{Reasons for being unaware of the misinformation.}
	\label{tab:reason}
	{
		\begin{tabular}{@{}lr@{}}
			\toprule
			Reason                  & Percentage \\ \midrule
			Inattention to elements related to misinformation   & 45.88\% \\
			Excessive trust in text                             & 15.74\% \\
			Wrong interpretation of visualizations              & 11.64\% \\
			Lacking Prior Knowledge                             & 10.24\%  \\
			Overlooking violation of visualization conventions  & 8.20\% \\
			Misunderstanding the reference relationship         & 5.48\% \\
			Lacking intention to find misinformation            & 2.73\% \\
			\bottomrule
		\end{tabular}
	}
\end{table}

\section{Discussion}
% \todo{impact}
% \review{
% R4: One would expect the readers to be interested in employing ways to counteract miscommunication. But what particular type of linking is used depends on how the *designer* decides to design the visualization solution. So I am confused by how this research is supposed to help counteract miscommunication. Even the idea proposed in Section 6.1 to find solutions to guide the reader's attention to misleading parts seems a bit far fetched to me. If one is able to detect misinformation why not make it explicit to the reader rather than diverting attention to it?
% }

% \todo{future work}
% \review{
% R3: It seems logical to me that in the text condition without linking the readers may very well have not inspected the visualization at all or only in a cursory way. In fact, from the study we are left with doubts to what is the main cause behind the results. Are the readers misled more in the text condition because they do not even look at the visualization or because there is no direct linking? We don't know ... but knowing that seems very relevant to me.
% }

% \todo{compare awareness with perceived credibility}
%The results of our experiment suggest a possible ``defensive deisgn''\cite{correll2017black} to combat misinformation in narrative visualization.

\subsection{Combat Misinformation from Narrative Visualization}
In this paper, we first explore the possible ways that narrative visualization can bring misinformation in its production and consumption process.
As suggested by Correll and Heer\cite{correll2017black}, ``we should enumerate the ways that visualizations can deceive or mislead.''.
Our work is a first step towards analyzing the space of misinformation in illustrated data stories.
Some forms of misinformation we have summarized, e.g., deceptive visualization\cite{pandey2015deceptive,lauer2020people} and text-visualization misalignment\cite{kong2019trust,kong2018frames}, have been studied in prior literature, and their threats to general audiences have been verified.
However, other forms of misinformation within our analysis have received relatively less attention this far.
For example, manipulating the layout in the Arranging step or organizing the story pieces' sequence in the Scripting step might result in readers' wrong decoding of visual information.
While there exist studies aiming at understanding the influence of layout or sequence on narrative visualization\cite{hullman2013deeper,bach2018design}, few of them concern misinformation.
Hence, more works are needed to deepen our understanding of misinformation in data storytelling.

The results of our crowdsourcing study suggest that the two common-used design methods, \textit{Annotation} and \textit{Linking}, can raise participants' awareness of misinformation.
We also found these two design methods can affect the perceived credibility of text and visualizations.
\revision{
However, the effects observed in our study look limited as most participants cannot find misinformation in the given data stories.
But, we believe the two methods, \textit{Annotation} and \textit{Linking}, have the potential to collaborate with other methods to form a defensive design to protect lay people from misinformation.
\textit{Annotation} and \textit{Linking} serve for bridging the perceptual gap between the text and visualization, which induce readers to pay attention to data-related narratives.
By adopting other attention-guiding methods to nudge readers reading data stories comprehensively and skeptically (e.g., explicit warning~\cite{law2020causal}), we can have solutions to ``Inattention to elements related misinformation'', the most important cause identified in our study for being unaware of misinformation. 
% For example, one can build a model to recognize misinformation in data stories. 
% For stories that the model has a certain level of confidence to be misinformation, explicit warnings can be added to the data story, and also the data story can be automatically transformed into the one with \textit{Annotation} or \textit{Linking}.
% Note that there already exist some technologies to automatically ``annotate''~\cite{lai2020automatic} and ``link''~\cite{kong2014extracting} narrative visualizations.

Worth mentioning, the two defensive design methods we chose as conditions in the experiment (\textit{Annotation} and \textit{Linking}) can be automatically applied to many data stories using existing technologies~\cite{lai2020automatic, kong2014extracting}.
With that said, we envision a possible future to defend misinformation: 
story or news release platforms, such as New York Times, can automatically transform the uploaded data stories from authors into a form that contains some defensive design methods and then release the stories to the public.
In that sense, the platform can establish a protective barrier in content dissemination from authors to readers. Even if advanced algorithms cannot detect misinformation, readers will still have the opportunity to be aware of misinformation.
}

\subsection{Design Implications}
% Despite the encouraging findings above, the majority of participants who experience strengthened connection between text and visualization in our study still failed to detect the false or inaccurate messages embedded in the given stories.
% To further improve the awareness of misinformation, besides considering other techniques such as word-scale visualization\cite{goffin2020interaction} to boost the text-visualization integration, 
In the crowdsourcing study, our qualitative analysis of participants' feedback provides insights into several design implications to mitigate misinformation in narrative visualization, as presented below.
% There are at least two ways: 
% (1) design stronger text-visualization integration using techniques like word-scale visualization\cite{goffin2020interaction}.
% (2) incorporating our methods with other techniques 
% rephrase
% Moreover, our qualitative analysis provides more insights into such a way to combat misinformation.
%We offer our three design thoughts motivated by the qualitative research below:

\textbf{Visual Design to Guide Attention}
Most participants in our experiment attribute their failure to find misinformation to their inattention (Table~\ref{tab:reason}).
Pennycook et al. \cite{pennycook2021shifting} postulate that directing people's attention to the accuracy of news can be a promising way to combat misinformation.
Consistent with this insights, our results indicate that a promising design direction is to lead readers to attend to the part(s) of stories where misinformation is prone to appear. %the reader's attention to the place where misinformation is prone to appear.%the importance of leading readers to attend to the part(s) of text and visualization that might affect information authenticity.
%Hence, a design direction worth exploring is to direct the reader's attention to the place where misinformation is prone to appear.
Such designs could potentially be a cure to ``overlooking violation of visualization conventions'' and ``misinterpreting the visualization''.
% two other reasons why they get misled mentioned by our participants.

% Visualint \cite{hopkins2020visualint} which highlights chart construction errors via sketches in situ is an excellent example of design techniques that call attention to places affecting information authenticity.
% However, as Visualint only concerns charts, more investigations involving text and other modalities of information are required in this design direction.

\textbf{Data-Friendly Text}
Text is an important component of narrative visualization, and researchers should rethink how its presentation can be improved to mitigate misinformation.
Our qualitative analysis reveals that text might cause misperception in readers, thereby leaving problematic messages in the stories unnoticed (e.g., ``excessive trust in text'' and ``misunderstanding the reference relationship'' in Table~\ref{tab:reason}).
Empirical evidence implies that whether textual description is closely linked to the data can affect reader understanding of the data story\cite{micallef2012assessing}.
A data-oriented narrative style of text, such as emphasizing the data source, clearly expressing the analysis method of the data, and more accurately specifying the visual elements of the discussion, may facilitate readers' judgment of the facts\cite{nenty2009writing}.
% Additionally, adding ``perspective sentence(s)'' that leverages ratios, ranks, or unit changes to provide contexts around numbers can help improve people’s ability to make sense of unfamiliar measurements\cite{riederer2018put}.
% Even though authors who deliberately want to spread misinformation may not write in this way, one might introduce automated text correction tools to make the text in the narrative visualization more data-friendly.
% Our quantitative and qualitative analysis reveals that the imbalance between people's trust in the text and the chart is an obstacle for them to find misinformation.
% % TODO not reasonable now
% A promising method to solve this might be to embrace visualization-friendly text in data storytelling to balance the trust to text and visualization.
% Micallef et. al\cite{micallef2012assessing} found the text removing number helps with bayesian reasoning tasks, which prove the impact of text format.
% However, few studies in narrative visualization related to the role of text format to misinformation.
% What's more, better writing of text might solve the problem "Misunderstand the reference relationship", which is listed in our results.
% For example, instead of using text to point to a large area of the chart, it is better to point to a data point more precisely.
% Researchers may be able to develop automated text correction tools to make the text in the narrative visualization more visualization-friendly.

\textbf{Knowledge-Augmented Visualization}
Some participants believe that the reason why they do not discover misinformation is a lack of understanding of visualization or the background of the story (``Lacking Prior Knowledge'' in Table~\ref{tab:reason}).
A feasible way to address this problem is to augment the original narrative visualization with external knowledge.
% to reduce the neglect of misinformation caused by the lack of necessary background. 
Similar ideas have been applied to improve awareness of diverse social opinions\cite{gao2018burst}.

% While the majority of participants with enhanced text-visualization integration in our study still do not aware the misinformation, there are at least two ways to further improve the awareness of misinformation: 
% (1) design stronger text-visualization integration using techniques like word-scale visualization\cite{goffin2020interaction}.
% (2) incorporating our methods with techniques mentioned above like visual design to guide attention.

\subsection{Limitation and Future Work}
Our study bears several limitations.
\revision{
First, we conducted experiments on three types of misinformation covering only two of the many causes of misinformation we identified in Table~\ref{tab:misinformation} (\textit{Break Conventions} and \textit{Text-visualization misalignment}). Our study material is only concerned with two visualization types (i.e., bar chart and line chart).
It is worth conducting experiments to understand what defensive designs can be effective to other causes of misinformation such as \textit{Manipulating order} and \textit{Obfuscation} as well as narrative visualizations with more diverse visualization types in the future.
Second, we did not randomize the type of misinformation for each story in the experiment to prevent the misinformation from being too obvious to see from the common knowledge.
Such design, however, can affect the effectiveness of our analysis in comparing different kinds of misinformation but not the conclusion about the two defensive designs, which is the focus of this paper.
Third, our measurements to participants' awareness of misinformation can be improved. 
We analyzed their awareness based on their reflection after reading stories.
More accurate measurements can be derived from other data collecting methods such as eye-tracking.}
Last but not least, we found \textit{Annotation} and \textit{Linking} do have effects, but we do not gain insights into how they work from our data. We hypothesize that they work by affecting readers' attention, which requires more investigation.
In the future, we will experiment with adopting instruments such as implicit measurement methods in psychology~\cite{karpinski2001attitudes} to obtain more accurate assessment of readers’ awareness of misinformation. We will further test readers' recall of the story and their potential attitude shift to discern the extent of impact of awareness.

\revision{
\section{Conclusion}
We introduce how misinformation can be injected into narrative visualization and explore possible defensive design methods to combat misinformation.
Misinformation has been ever dangerous to human society with the ability of online social media to spread information quickly.
Combating misinformation is an essential but also challenging task.
In this paper, we choose two commonly used methods in narrative visualization as candidates of defensive designs.
Our study results show the potential of these two and other similar methods to be a part of the force to combat misinformation.
Also, the number of people who do not find misinformation in our study again let us notice the challenge of fighting an infodemic.
}
\acknowledgments{
We thank Yujie Zheng for valuable input.
We also thank the anonymous reviewers for their feedback and comments that helped improve this paper.
}

\bibliographystyle{abbrv-doi}

\bibliography{template}

\begin{thebibliography}{10}

\bibitem{andreou_2020}
A.~Andreou.
\newblock Comparing the gov't covid19 maps for end of september and start of
  october, it looks as if things are getting better.until you notice they've
  changed the numbers corresponding to each colour. had they used the same
  ones, most of the country would be red or dark red.\#bitsneaky
  pic.twitter.com/61u1dd4g1u.
\newblock \url{https://twitter.com/sturdyAlex/status/1314563677144657922}, Oct
  2020.
\newblock Accessed: March 31, 2021.

\bibitem{bach2018design}
B.~Bach, Z.~Wang, M.~Farinella, D.~Murray-Rust, and N.~Henry~Riche.
\newblock Design patterns for data comics.
\newblock In {\em Proceedings of the ACM Conference on Human Factors in
  Computing Systems}, pp. 1--12, 2018.

\bibitem{borgo2018information}
R.~Borgo, L.~Micallef, B.~Bach, F.~McGee, and B.~Lee.
\newblock Information visualization evaluation using crowdsourcing.
\newblock In {\em Computer Graphics Forum}, vol.~37, pp. 573--595. Wiley Online
  Library, 2018.

\bibitem{braun2012thematic}
V.~Braun and V.~Clarke.
\newblock Thematic analysis.
\newblock 2012.

\bibitem{bryan2016temporal}
C.~Bryan, K.-L. Ma, and J.~Woodring.
\newblock Temporal summary images: An approach to narrative visualization via
  interactive annotation generation and placement.
\newblock {\em IEEE transactions on visualization and computer graphics},
  23(1):511--520, 2016.

\bibitem{canham2010effects}
M.~Canham and M.~Hegarty.
\newblock Effects of knowledge and display design on comprehension of complex
  graphics.
\newblock {\em Learning and Instruction}, 20(2):155--166, 2010.

\bibitem{chen2021vizlinter}
Q.~Chen, F.~Sun, X.~Xu, Z.~Chen, J.~Wang, and N.~Cao.
\newblock Vizlinter: A linter and fixer framework for data visualization.
\newblock {\em IEEE transactions on visualization and computer graphics}, 2021.

\bibitem{chen2018supporting}
S.~Chen, J.~Li, G.~Andrienko, N.~Andrienko, Y.~Wang, P.~H. Nguyen, and
  C.~Turkay.
\newblock Supporting story synthesis: Bridging the gap between visual analytics
  and storytelling.
\newblock {\em IEEE Transactions on Visualization and Computer Graphics},
  26(7):2499--2516, 2018.

\bibitem{chou2018addressing}
W.-Y.~S. Chou, A.~Oh, and W.~M. Klein.
\newblock Addressing health-related misinformation on social media.
\newblock {\em Jama}, 320(23):2417--2418, 2018.

\bibitem{cockburn2018hark}
A.~Cockburn, C.~Gutwin, and A.~Dix.
\newblock Hark no more: on the preregistration of chi experiments.
\newblock In {\em Proceedings of the ACM Conference on Human Factors in
  Computing Systems}, pp. 1--12, 2018.

\bibitem{cook2018deconstructing}
J.~Cook, P.~Ellerton, and D.~Kinkead.
\newblock Deconstructing climate misinformation to identify reasoning errors.
\newblock {\em Environmental Research Letters}, 13(2):024018, Feb. 2018.

\bibitem{correll2016surprise}
M.~Correll and J.~Heer.
\newblock Surprise! bayesian weighting for de-biasing thematic maps.
\newblock {\em IEEE Transactions on Visualization and Computer Graphics},
  23(1):651--660, 2016.

\bibitem{correll2017black}
M.~Correll and J.~Heer.
\newblock Black hat visualization.
\newblock In {\em Workshop on Dealing with Cognitive Biases in Visualisations
  (DECISIVe), IEEE VIS}, 2017.

\bibitem{doan2021misrepresenting}
S.~Doan.
\newblock Misrepresenting covid-19: Lying with charts during the second golden
  age of data design.
\newblock {\em Journal of Business and Technical Communication}, 35(1):73--79,
  2021.

\bibitem{ecker2014correcting}
U.~K. Ecker, B.~Swire, and S.~Lewandowsky.
\newblock Correcting misinformation—a challenge for education and cognitive
  science.
\newblock 2014.

\bibitem{feenberg2017s}
D.~Feenberg, I.~Ganguli, P.~Gaule, and J.~Gruber.
\newblock It’s good to be first: Order bias in reading and citing nber
  working papers.
\newblock {\em Review of Economics and Statistics}, 99(1):32--39, 2017.

\bibitem{friel2001making}
S.~N. Friel, F.~R. Curcio, and G.~W. Bright.
\newblock Making sense of graphs: Critical factors influencing comprehension
  and instructional implications.
\newblock {\em Journal for Research in Mathematics Education}, 32(2):124--158,
  2001.

\bibitem{gao2018burst}
M.~Gao, H.~J. Do, and W.-T. Fu.
\newblock Burst your bubble! an intelligent system for improving awareness of
  diverse social opinions.
\newblock In {\em 23rd International Conference on Intelligent User
  Interfaces}, pp. 371--383, 2018.

\bibitem{garrett2014implications}
R.~K. Garrett, S.~D. Gvirsman, B.~K. Johnson, Y.~Tsfati, R.~Neo, and A.~Dal.
\newblock Implications of pro-and counterattitudinal information exposure for
  affective polarization.
\newblock {\em Human Communication Research}, 40(3):309--332, 2014.

\bibitem{guo2019future}
B.~Guo, Y.~Ding, L.~Yao, Y.~Liang, and Z.~Yu.
\newblock The future of false information detection on social media: New
  perspectives and trends.
\newblock {\em ACM Comput. Surv.}, 53(4), July 2020. doi: {{%
10\hspace{.1pt}\discretionary{.}{%
}{.}\hspace{.4pt}1145\discretionary{/}{%
}{/}3393880}}


\bibitem{guo2017you}
Y.~Guo, C.~Binnig, and T.~Kraska.
\newblock What you see is not what you get! detecting simpson's paradoxes
  during data exploration.
\newblock In {\em Proceedings of the 2nd Workshop on Human-In-the-Loop Data
  Analytics}, pp. 1--5, 2017.

\bibitem{6064988}
J.~{Hullman} and N.~{Diakopoulos}.
\newblock Visualization rhetoric: Framing effects in narrative visualization.
\newblock {\em IEEE Transactions on Visualization and Computer Graphics},
  17(12):2231--2240, 2011. doi: {{%
10\hspace{.1pt}\discretionary{.}{%
}{.}\hspace{.4pt}1109\discretionary{/}{%
}{/}TVCG\hspace{.1pt}\discretionary{.}{%
}{.}\hspace{.4pt}2011\hspace{.1pt}\discretionary{.}{%
}{.}\hspace{.4pt}255}}


\bibitem{hullman2013deeper}
J.~Hullman, S.~Drucker, N.~H. Riche, B.~Lee, D.~Fisher, and E.~Adar.
\newblock A deeper understanding of sequence in narrative visualization.
\newblock {\em IEEE Transactions on Visualization and Computer Graphics},
  19(12):2406--2415, 2013.

\bibitem{johnston1984prior}
P.~Johnston.
\newblock Prior knowledge and reading comprehension test bias.
\newblock {\em Reading Research Quarterly}, pp. 219--239, 1984.

\bibitem{karpinski2001attitudes}
A.~Karpinski and J.~L. Hilton.
\newblock Attitudes and the implicit association test.
\newblock {\em Journal of Personality and Social Psychology}, 81(5):774, 2001.

\bibitem{kong2018frames}
H.-K. Kong, Z.~Liu, and K.~Karahalios.
\newblock Frames and slants in titles of visualizations on controversial
  topics.
\newblock In {\em Proceedings of the ACM Conference on Human Factors in
  Computing Systems}, pp. 1--12, 2018.

\bibitem{kong2019trust}
H.-K. Kong, Z.~Liu, and K.~Karahalios.
\newblock Trust and recall of information across varying degrees of
  title-visualization misalignment.
\newblock In {\em Proceedings of the ACM Conference on Human Factors in
  Computing Systems}, pp. 1--13, 2019.

\bibitem{kong2014extracting}
N.~Kong, M.~A. Hearst, and M.~Agrawala.
\newblock Extracting references between text and charts via crowdsourcing.
\newblock In {\em Proceedings of the SIGCHI conference on Human Factors in
  Computing Systems}, pp. 31--40, 2014.

\bibitem{kwon2014visjockey}
B.~C. Kwon, F.~Stoffel, D.~J{\"a}ckle, B.~Lee, and D.~Keim.
\newblock Visjockey: Enriching data stories through orchestrated interactive
  visualization.
\newblock In {\em Poster compendium of the computation+ journalism symposium},
  vol.~3, p.~3, 2014.

\bibitem{lai2020automatic}
C.~Lai, Z.~Lin, R.~Jiang, Y.~Han, C.~Liu, and X.~Yuan.
\newblock Automatic annotation synchronizing with textual description for
  visualization.
\newblock In {\em Proceedings of the ACM Conference on Human Factors in
  Computing Systems}, pp. 1--13, 2020.

\bibitem{lauer2020people}
C.~Lauer and S.~O'Brien.
\newblock How people are influenced by deceptive tactics in everyday charts and
  graphs.
\newblock {\em IEEE Transactions on Professional Communication},
  63(4):327--340, 2020.

\bibitem{law2020causal}
P.-M. Law, L.~Y.-H. Lo, A.~Endert, J.~Stasko, and H.~Qu.
\newblock Causal perception in question-answering systems.
\newblock In {\em Proceedings of the 2021 CHI Conference on Human Factors in
  Computing Systems}, pp. 1--15, 2021.

\bibitem{7274435}
B.~{Lee}, N.~H. {Riche}, P.~{Isenberg}, and S.~{Carpendale}.
\newblock More than telling a story: Transforming data into visually shared
  stories.
\newblock {\em IEEE Computer Graphics and Applications}, 35(5):84--90, 2015.
  doi: {{%
10\hspace{.1pt}\discretionary{.}{%
}{.}\hspace{.4pt}1109\discretionary{/}{%
}{/}MCG\hspace{.1pt}\discretionary{.}{%
}{.}\hspace{.4pt}2015\hspace{.1pt}\discretionary{.}{%
}{.}\hspace{.4pt}99}}


\bibitem{lee2021viral}
C.~Lee, T.~Yang, G.~Inchoco, G.~M. Jones, and A.~Satyanarayan.
\newblock Viral visualizations: How coronavirus skeptics use orthodox data
  practices to promote unorthodox science online.
\newblock {\em arXiv preprint arXiv:2101.07993}, 2021.

\bibitem{matute2015illusions}
H.~Matute, F.~Blanco, I.~Yarritu, M.~D{\'\i}az-Lago, M.~A. Vadillo, and
  I.~Barberia.
\newblock Illusions of causality: how they bias our everyday thinking and how
  they could be reduced.
\newblock {\em Frontiers in Psychology}, 6:888, 2015.

\bibitem{mckenna2017visual}
S.~McKenna, N.~Henry~Riche, B.~Lee, J.~Boy, and M.~Meyer.
\newblock Visual narrative flow: Exploring factors shaping data visualization
  story reading experiences.
\newblock In {\em Computer Graphics Forum}, vol.~36, pp. 377--387. Wiley Online
  Library, 2017.

\bibitem{10.1145/3313831.3376420}
A.~McNutt, G.~Kindlmann, and M.~Correll.
\newblock Surfacing visualization mirages.
\newblock In {\em Proceedings of the ACM Conference on Human Factors in
  Computing Systems}, p. 1–16, 2020. doi: {{%
10\hspace{.1pt}\discretionary{.}{%
}{.}\hspace{.4pt}1145\discretionary{/}{%
}{/}3313831\hspace{.1pt}\discretionary{.}{%
}{.}\hspace{.4pt}3376420}}


\bibitem{meirick2013motivated}
P.~C. Meirick.
\newblock Motivated misperception? party, education, partisan news, and belief
  in “death panels”.
\newblock {\em Journalism \& Mass Communication Quarterly}, 90(1):39--57, 2013.

\bibitem{micallef2012assessing}
L.~Micallef, P.~Dragicevic, and J.-D. Fekete.
\newblock Assessing the effect of visualizations on bayesian reasoning through
  crowdsourcing.
\newblock {\em IEEE Transactions on Visualization and Computer Graphics},
  18(12):2536--2545, 2012.

\bibitem{nasa_2021}
NASA.
\newblock Global surface temperature.
\newblock \url{https://climate.nasa.gov/vital-signs/global-temperature}.
\newblock Accessed: March 31, 2021.

\bibitem{nenty2009writing}
H.~J. Nenty.
\newblock Writing a quantitative research thesis.
\newblock {\em International Journal of Educational Sciences}, 1(1):19--32,
  2009.

\bibitem{network_2020}
T.~L. Network.
\newblock What's going on in this graph? | immigration shifts.
\newblock
  \url{https://www.nytimes.com/2020/04/30/learning/whats-going-on-in-this-graph-immigration-shifts.html},
  Apr. 2020.
\newblock Accessed: March 31, 2021.

\bibitem{newman2018effects}
A.~Newman, Z.~Bylinskii, S.~Haroz, S.~Madan, F.~Durand, and A.~Oliva.
\newblock Effects of title wording on memory of trends in line graphs.
\newblock {\em Journal of Vision}, 18(10):837--837, 2018.

\bibitem{6876023}
A.~V. {Pandey}, A.~{Manivannan}, O.~{Nov}, M.~{Satterthwaite}, and
  E.~{Bertini}.
\newblock The persuasive power of data visualization.
\newblock {\em IEEE Transactions on Visualization and Computer Graphics},
  20(12):2211--2220, 2014. doi: {{%
10\hspace{.1pt}\discretionary{.}{%
}{.}\hspace{.4pt}1109\discretionary{/}{%
}{/}TVCG\hspace{.1pt}\discretionary{.}{%
}{.}\hspace{.4pt}2014\hspace{.1pt}\discretionary{.}{%
}{.}\hspace{.4pt}2346419}}


\bibitem{pandey2015deceptive}
A.~V. Pandey, K.~Rall, M.~L. Satterthwaite, O.~Nov, and E.~Bertini.
\newblock How deceptive are deceptive visualizations? an empirical analysis of
  common distortion techniques.
\newblock In {\em Proceedings of the ACM Conference on Human Factors in
  Computing Systems}, pp. 1469--1478, 2015.

\bibitem{pennycook2021shifting}
G.~Pennycook, Z.~Epstein, M.~Mosleh, A.~A. Arechar, D.~Eckles, and D.~G. Rand.
\newblock Shifting attention to accuracy can reduce misinformation online.
\newblock {\em Nature}, pp. 1--6, 2021.

\bibitem{petty2012communication}
R.~E. Petty and J.~T. Cacioppo.
\newblock {\em Communication and persuasion: Central and peripheral routes to
  attitude change}.
\newblock Springer Science \& Business Media, 2012.

\bibitem{wp_a}
T.~W. Post.
\newblock 75 years of major refugee crises around the world.
\newblock
  \url{https://www.washingtonpost.com/graphics/world/historical-migrant-crisis}.
\newblock Accessed: January 15, 2021.

\bibitem{roozenbeek2020susceptibility}
J.~Roozenbeek, C.~R. Schneider, S.~Dryhurst, J.~Kerr, A.~L. Freeman,
  G.~Recchia, A.~M. Van Der~Bles, and S.~Van Der~Linden.
\newblock Susceptibility to misinformation about covid-19 around the world.
\newblock {\em Royal Society Open Science}, 7(10):201199, 2020.

\bibitem{sacha2015role}
D.~Sacha, H.~Senaratne, B.~C. Kwon, G.~Ellis, and D.~A. Keim.
\newblock The role of uncertainty, awareness, and trust in visual analytics.
\newblock {\em IEEE Transactions on Visualization and Computer Graphics},
  22(1):240--249, 2015.

\bibitem{segel2010narrative}
E.~Segel and J.~Heer.
\newblock Narrative visualization: Telling stories with data.
\newblock {\em IEEE Transactions on Visualization and Computer Graphics},
  16(6):1139--1148, 2010.

\bibitem{shi2020calliope}
D.~Shi, X.~Xu, F.~Sun, Y.~Shi, and N.~Cao.
\newblock Calliope: Automatic visual data story generation from a spreadsheet.
\newblock {\em IEEE Transactions on Visualization and Computer Graphics}, 2020.

\bibitem{stevenson2010oxford}
A.~Stevenson.
\newblock {\em Oxford dictionary of English}.
\newblock Oxford University Press, USA, 2010.

\bibitem{8585669}
E.~{Wall}, L.~M. {Blaha}, L.~{Franklin}, and A.~{Endert}.
\newblock Warning, bias may occur: A proposed approach to detecting cognitive
  bias in interactive visual analytics.
\newblock In {\em IEEE Conference on Visual Analytics Science and Technology
  (VAST)}, pp. 104--115, 2017. doi: {{%
10\hspace{.1pt}\discretionary{.}{%
}{.}\hspace{.4pt}1109\discretionary{/}{%
}{/}VAST\hspace{.1pt}\discretionary{.}{%
}{.}\hspace{.4pt}2017\hspace{.1pt}\discretionary{.}{%
}{.}\hspace{.4pt}8585669}}


\bibitem{wang2019comparing}
Z.~Wang, S.~Wang, M.~Farinella, D.~Murray-Rust, N.~Henry~Riche, and B.~Bach.
\newblock Comparing effectiveness and engagement of data comics and
  infographics.
\newblock In {\em Proceedings of the ACM Conference on Human Factors in
  Computing Systems}, pp. 1--12, 2019.

\bibitem{wu2019misinformation}
L.~Wu, F.~Morstatter, K.~M. Carley, and H.~Liu.
\newblock Misinformation in social media: definition, manipulation, and
  detection.
\newblock {\em ACM SIGKDD Explorations Newsletter}, 21(2):80--90, 2019.

\bibitem{zhi2019linking}
Q.~Zhi, A.~Ottley, and R.~Metoyer.
\newblock Linking and layout: Exploring the integration of text and
  visualization in storytelling.
\newblock In {\em Computer Graphics Forum}, vol.~38, pp. 675--685. Wiley Online
  Library, 2019.

\end{thebibliography}
\end{document}